# Manipulating non-intrinsic outputs of a non-Hermitian coupled system by weak external driving

He Feng[1#], Yi-Fan Xiang[1#], Wei Li[1*], and Xue-Hua Wang[1*]

[1]State Key Laboratory of Optoelectronic Materials and Technologies, School of Physics, Sun Yat-sen University, Guangzhou 510275, China

[*]Corresponding author(s). E-mail(s): liwei373@mail.sysu.edu.cn; wangxueh@mail.sysu.edu.cn

[#]These authors contributed equally to this work

## Abstract

Measurement provides access to exploring and understanding nature and finds widespread applications. It is generally believed that the weak external driving used in measurement introduces negligible disturbance to the intrinsic output of the probed system. Here, we reveal that dissipation in a non-Hermitian quantum coupled system induces interference between self-excitation and coupling-excitation spectral functions, giving rise to non-intrinsic outputs. These non-intrinsic outputs can be effectively manipulated by the ratio of weak external driving strengths applied to two channels, leading to significant phenomena: extreme suppression of dissipation by convergence effect of high-loss mode towards low-loss mode, and giant shift of resonant frequency instead of original spectral Rabi splitting. Crucially, we discover the Pythagorean relation linking eigenlevel splitting, spectral Rabi splitting and average total decay rate, enabling precise measurement of the eigenlevel splitting and coupling strength. Numerical simulations confirm that the interference outputs also exist in classical coupled systems. Our work provides a profound insight into measurement and non-Hermitian physics.

## Introduction

Measurement is central to exploring and understanding nature and underpins progress across modern science and advanced technologies[1,2]. Among its diverse forms, spectroscopy has long served as one of the most powerful and versatile tools for probing quantum systems[3,4], enabling access to eigenlevel structures, interaction mechanisms, and system-environment couplings across multiple platforms ranging from atomic and molecular systems to solid-state, photonic, and hybrid quantum architectures. In idealized isolated or weakly dissipative systems under weak external driving, the measurement functions as passive probes that do not significantly alter a system's intrinsic properties. The output spectra arise from the system's linear response to weak external perturbations and truly map the eigenlevel structure and intrinsic dynamics of the probed systems. Consequently, these spectra are generally regarded as an intrinsic fingerprint of the eigenstates in the probed quantum systems[5-7]. However, in realistic open quantum systems—especially in non-Hermitian coupled systems—where dissipation and decoherence are unavoidable and significant. Whether this conventional understanding remains valid in such context represents a fundamental question of profound physical significance.

Non-Hermitian physics, emerging as a powerful paradigm for describing the dissipative dynamics of open quantum systems, has attracted tremendous research interest in recent years[8]. Its core principles have been successfully implemented across a broad range of platforms, including ultracold atoms[9,10], acoustics[11-13], metamaterials[14-16], and photonics[17-20]. In these platforms, non-Hermitian systems give rise to a host of nontrivial phenomena, such as the non-Hermitian skin effect[21-23], nonreciprocity[24-26], bound states in the continuum (BICs)[27,28], room-temperature quantum strong coupling[29-32], exceptional points (EPs)[33-35] and the associated parity-time symmetry breaking[36-38], topological phase transitions[39] as well as many other nontrivial topological features[40-42]. Undoubtedly, these cutting-edge explorations have significantly advanced the development of applications in lasers[43,44], high-sensitivity sensors[45-47], electrical circuits[48,49], and on-chip quantum devices[50,51]. Despite intense research activities and remarkable advances in non-Hermitian systems, the above-outlined fundamental question remains unresolved. It remains widely believed that weak external driving only modulates the intensity of output spectral signals without exerting observable influence on the spectral structure of the probed systems itself.

In this paper, we reveal that weak external driving applied to the each channel (i.e., each mode) of a non-Hermitian quantum coupled system induces the self-excitation and coupling-excitation spectral functions. Dissipation causes them to interfere with each other, thereby generating non-intrinsic outputs for each channel that no longer represent the intrinsic fingerprints of the eigenstates in the non-Hermitian coupled system. These non-intrinsic outputs can be tailored on-demand by the driving strength ratio applied to two channels. This externally controllable interference triggers a set of new physical phenomena, including a thousand-fold dissipation suppression enabled by convergence effect of high-loss mode toward low-loss mode under very weak coupling, and a giant shift of resonant frequency—instead of the original spectral Rabi splitting (SRS)—under strong coupling. Importantly, we find a Pythagorean relationship that connects the eigenlevel splitting (LS), the SRS and average total decay rate, providing a new paradigm for precisely determining the intrinsic LS and coupling strength of non-Hermitian coupled systems from experimental measurements. Finally, we verify the predicted spectral behaviors via numerical simulations in a photonic-plasmonic hybrid system, demonstrating that the excitation interference is universal across both classic and quantum coupled systems. It is worth pointing out that when the dissipation vanishes or is negligible, this dissipation-induced interference disappears and consequently the conventional weak measurement theory becomes valid. Our work opens new opportunities for programmable measurement, controllable spectral engineering, and mode loss suppression in non-Hermitian coupled systems.

## Theory

A non-Hermitian system coupled by two modes (i.e., two subsystems) is described by the following effective Hamiltonian[29,52]

$$H_{\mathrm{sys}} = (\omega_d - i\Gamma_d/2)d^\dagger d + (\omega_c - i\Gamma_c/2)c^\dagger c + g(dc^\dagger + cd^\dagger). \tag{1}$$

Here $d$ and $c$ ($d^\dagger$ and $c^\dagger$) are the annihilation (creation) operators of the two modes, $\omega_d$ ($\omega_c$) and $\Gamma_d$ ($\Gamma_c$) is their resonance frequency and dissipative (i.e., damping) linewidth, respectively; $g$ is the coupling strength between them. The eigenlevels of the coupled-system are given by

$$\omega_\pm = (\omega_d + \omega_c)/2 - i(\Gamma_d + \Gamma_c)/4 \pm \Delta_{\mathrm{LS}}/2, \tag{2}$$

where $\Delta_{\mathrm{LS}} = \sqrt{4g^2 + [\omega_d - \omega_c - i(\Gamma_d - \Gamma_c)/2]^2}$ is the LS. At resonance ($\omega_d = \omega_c$), the LS is expressed as $\Delta_{\mathrm{LS}} = 2\sqrt{g^2 - g_{\mathrm{EP}}^2}$ with $g_{\mathrm{EP}} = |\Gamma_d - \Gamma_c|/4$ being the critical coupling strength at the exceptional point (EP). For $g < g_{\mathrm{EP}}$, the coupled system resides in the weak-coupling regime with degenerate eigenlevels. When $g > g_{\mathrm{EP}}$, the eigenlevel splits into two branches[32], which is the precondition for the coupled system to enter the strong-coupling regime (Fig. 1a). For the sake of statement, we denote the two modes as channels $d$ and $c$ (Ch-$d$ and Ch-$c$) in the following text, respectively.

In order to probe the eigenlevels of the coupled system, a weak external driving as shown in Eq.(3) is generally applied to excite the system for outputting absorption or fluorescence spectra

$$H_{\mathrm{dri}} = E_d(d^\dagger + d) + E_c(c^\dagger + c), \tag{3}$$

where $E_d$ and $E_c$ are the external driving strengths applied to Ch-$d$ and Ch-$c$, respectively. Previous perturbation and non-perturbation schemes[53-57] demonstrate that the weak external driving has no observable influence on the output spectra of the probed systems except the spectral signal intensity. Therefore, the output spectra exhibit an intrinsic fingerprint-like correspondence with the probed system (Supplementary Section 1), although some differences exist between the LS and SRS[56,58].

Here, we directly solve the Heisenberg equations of motion from the total Hamiltonian $H_{\mathrm{total}} = H_{\mathrm{sys}} + H_{\mathrm{dri}}$ to obtain the spectral response of the probed system. Under weak driving and negligible nonlinear effects, the output spectra of the Ch-$d$, $S_d$, and the Ch-$c$, $S_c$, are analytically obtained by the Laplace transform method (Supplementary Section 2) as follows,

$$S_d \propto \left|E_d\psi_d + E_c\psi_g\right|^2, \tag{4a}$$

$$S_c \propto \left|E_c\psi_c + E_d\psi_g\right|^2, \tag{4b}$$

where $\psi_{d(c)} = (\omega - \omega_{c(d)} + i\Gamma_{c(d)}/2)/\chi$ and $\psi_g = g/\chi$ denote the channels' self-excitation and coupling-excitation spectral functions, respectively, with the $\chi = (\omega - \omega_+)(\omega - \omega_-)$. The total spectrum of the coupled system can be expressed as $S_{\mathrm{total}} \propto A_d S_d + A_c S_c$, where $A_d$ and $A_c$ are the corresponding weight factors of the two channels' spectra, respectively.

Equations (4a) and (4b) demonstrate that when a channel is driven, it excites spectral functions not only in itself but also in the other channel by the coupling. The output of each channel arises from the interference of the self-excitation and coupling-excitation spectral functions, and the total spectrum $S_{\mathrm{total}}$ follows as a linear superposition of the two channel-resolved spectra. They can be effectively manipulated by the ratio of external driving intensities applied to two channels, which fundamentally shapes the measured spectral response. As a result, the observed spectra are non-intrinsic and cease to be direct manifestations of the system's intrinsic eigenlevel structures; instead, they are jointly shaped by the system's eigenlevel structures, the driving intensity ratio, and the probing channels (Fig. 1b). This interference output transcends the conventional measurement theory that the weak external driving has no observable effect on the output of the probed system, offering a new degree of freedom to manipulate the output spectra without altering the system's internal architecture. It is worth pointing out that the interference effect stems from the dissipation, and it vanishes at $\Gamma_{c(d)} = 0$ since the spectral functions become purely real-valued. In addition, no interference occur under only the single-channel driving (Fig. S1 in Supplementary Section 2.3).

Existing experimental techniques already enable selective enhancement of the contribution from a specific channel in measured spectra, and even allow for full isolation of single-channel spectra[59]. This provides a readily accessible pathway to experimentally validate our theory, as well as the predictions derived from the external driving-induced interference that is mediated by the selection of driving and probing channels.

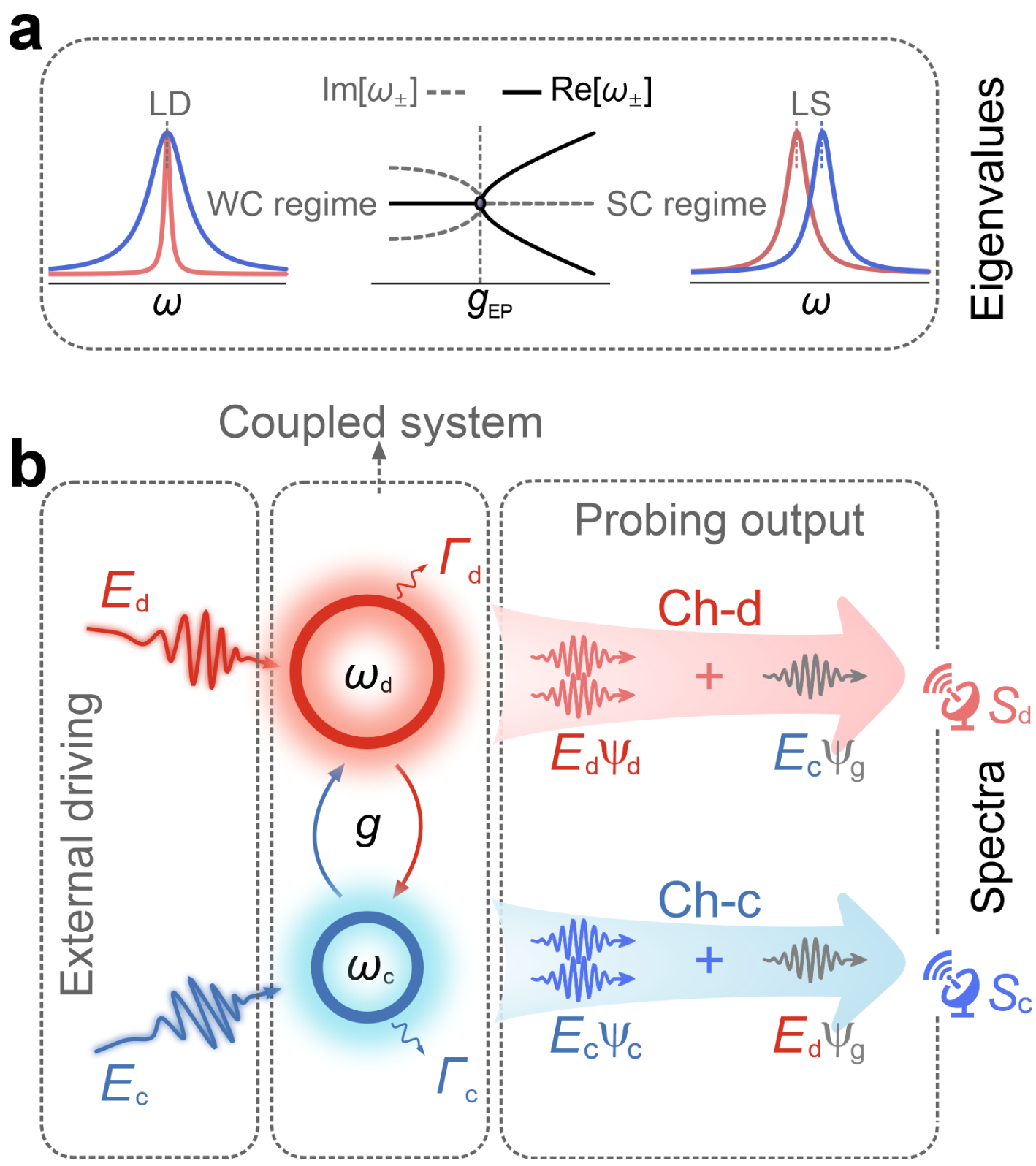

**Fig. 1 | Mechanism of spectral-eigenlevel deviation. a,** Intrinsic eigenlevel structure of the non-Hermitian coupled system. **b,** Determinants of the output spectra in the coupled system.

## On-demand output of non-intrinsic spectra

Weak external driving induces intra-channel interference between the channel's self-excitation and coupling-excitation spectral functions, giving rise to plentiful and diverse spectral responses across both channels and challenging the conventional view that spectra serve as fingerprints of intrinsic eigenstates. The on-demand manipulation of the output spectra at resonance by controlling the external driving configuration is shown in Fig. 2. Figure 2a illustrates the outputs of both channels in (i) the weak-coupling regime and (ii) the strong-coupling regime, demonstrating pronounced spectral variations as the driving intensity ratios $E_d\!:\!E_c$ applied to the two channels are adjusted.

In the weak-coupling regime, Fig. 2a(i) displays the characteristic spectra under three different driving intensity ratios. Notably, the weak external driving exerts a significant influence on the output spectrum of high-loss Ch-$d$, manifesting as distinct spectral features including a narrow single peak, a Fano lineshape, and asymmetric splitting. In contrast, the spectra of low-loss Ch-$c$ basically retain the narrow single-peak lineshape characteristic of the uncoupled bare mode, indicating that the impact on Ch-$c$ remains minor. This observation aligns with the conventional picture that weak driving does not substantially perturb the intrinsic output of a low-loss system. It suggests that the influence of weak external driving on the probed system can be amplified by increasing dissipation.

In the strong-coupling regime, the output spectra of both channels are presented in Fig. 2a(ii). It is evident that for a given driving intensity ratio, the spectra of Ch-$d$ and Ch-$c$ differ markedly, and across different ratios they appear to originate from entirely different coupling systems. Owing to interference between the self-excitation and coupling-excitation spectral functions, the spectral lineshapes exhibit obvious asymmetry, in vast contrast to the symmetry predicted by the conventional theory (Supplementary Section 3). The splitting width of the output spectrum from the predominantly driven channel exceeds that of the other channel. Interestingly, Fig. 2a(ii) shows a clear discrepancy between the SRS and LS marked by the dashed lines. Only when the driving on the low-loss Ch-$c$ ($g \gg \Gamma_c$) dominates, the SRS from Ch-$c$ is approximately equal to the LS of the coupled system. The modulation of the total system spectra is presented in Supplementary Section 4. Figure 2b shows the modulation of the coupled system's output spectra as the driving intensity ratio is continuously varied under different coupling strengths, enabling on-demand tuning of increasingly diverse spectral features. The modulation behavior of the output spectra with varying coupling strengths under different driving intensity ratios is further detailed in Supplementary Section 5.

These findings reveal that when both modes of a coupled system are simultaneously excited by weak external driving, the interference occurs between the self-excitation and coupling-excitation spectral functions. This interference effect depends strongly on the driving intensity ratio between the two channels, and can be further amplified by increasing dissipation. Consequently, tuning the driving intensity ratio enables the generation of diverse non-intrinsic spectral outputs, providing a flexible and efficient external degree of freedom for on-demand spectral output.

Notably, our theory unifies previously reported limiting cases in which only a single channel is driven[53-57] (further details are provided in Supplementary Section 6). The self-excitation and coupling-excitation spectra in Eq. (4) respectively correspond to the emission spectra from the emitter channel[53] and from the cavity channel[54] in the framework of cavity quantum electrodynamics. The total absorption spectrum of the coupled system comprising a plasmonic mode and quantum emitters, derived from our model, coincides with that obtained by the Zubarev Green's function method[55,56]. Microcavity-engineered plasmonic resonances[57] can also be faithfully reproduced within our theoretical framework. The self-excitation and coupling-excitation spectral fucntions, and their interference revealed in this work, provide foundational insight into non-Hermitian physics of coupled systems under weak external driving and lead to significant physical phenomena, thereby opening a route toward on-demand manipulation of system outputs.

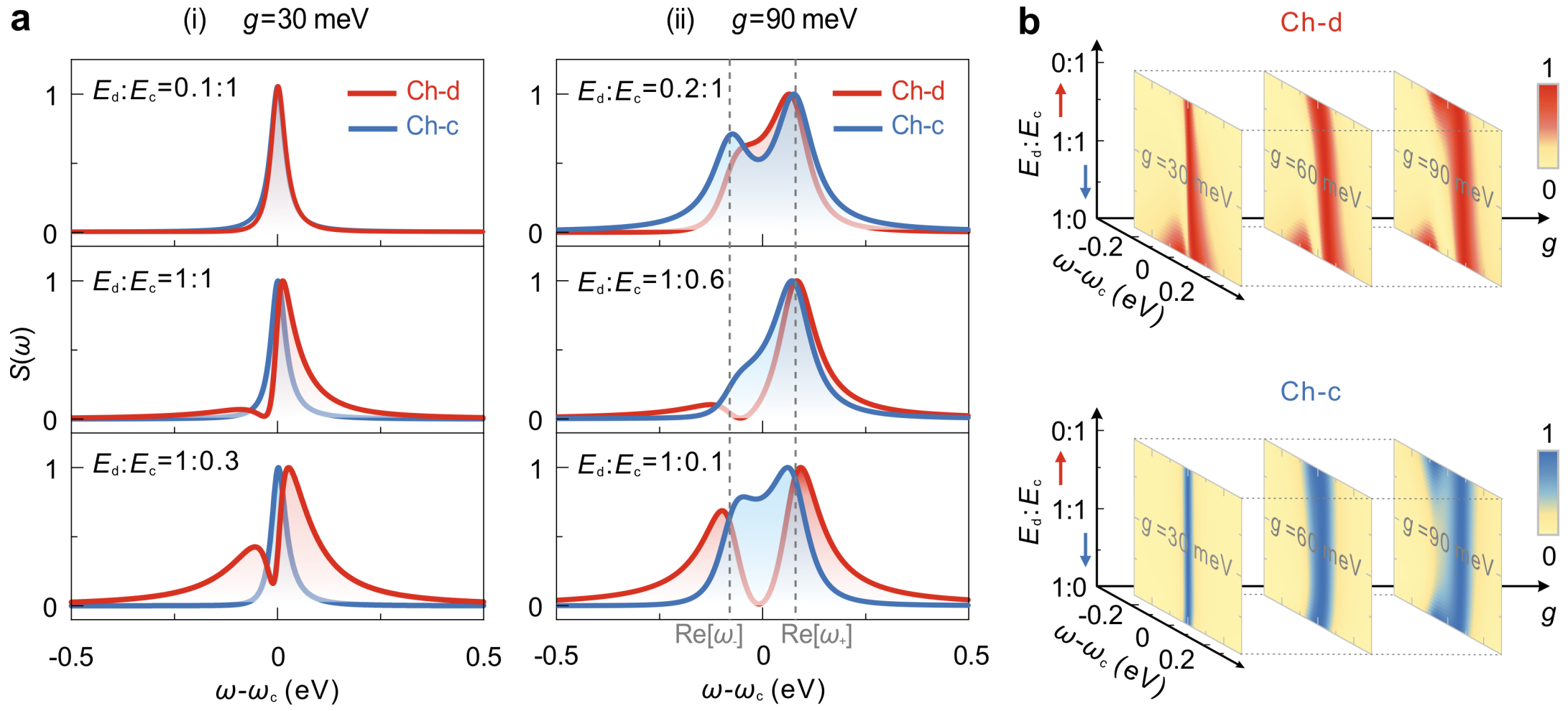

**Fig. 2 | On-demand spectral manipulation. a,** Normalized spectra from Ch-$d$ and Ch-$c$ under different $E_d$: $E_c$ at $g$ = 30 meV (i) and 90 meV (ii). The dashed lines in a(ii) mark the eigenvalues of the coupled system. **b,** Spectral evolution maps of Ch-$d$ and Ch-$c$ versus $E_d$: $E_c$ for $g$ = 30, 60, and 90 meV. In calculations, $\Gamma_d$ = 200 meV and $\Gamma_c$ = 20 meV.

## Convergence effect of high-loss mode towards low-loss mode

Mode dissipation signifies irreversible energy and information loss, phase decoherence, low quality factor, and related performance limitations. Its effective suppression is critical for energy storage, optical communication, quantum technology, nanoscale lasers, and single-molecule sensing. Within our unified framework, we demonstrate that by selectively driving the low-loss mode channel, the high loss of the other mode can be substantially suppressed.

As shown in Fig. 3a(i), when only Ch-$c$ is driven, its spectrum exhibits coupling-induced broadening, whereas the spectrum of Ch-$d$ exhibits a narrowing linewidth in the weak-coupling regime ($g$ = 45 meV). As $g$ decreases to 2 meV (very small), the linewidth of the Ch-$d$ output spectrum is strikingly compressed by approximately 100- and 1000-fold, as shown in Figs. 3a(ii) and 3a(iii). In other words, the quality factor of the mode $d$ is enhanced by two to three orders of magnitude, approaching that of the bare mode-$c$. Figure S6 further demonstrates that the total output spectrum of the coupled system also features an ultra-narrow linewidth (Supplementary Section 7). Figure 3b plots the evolution of the full width at half maximum (FWHM) for the spectra from Ch-$c$, Ch-$d$, and the total spectra as a function of $g$ at $\Gamma_c$ = 2 meV, under the exclusive driving of Ch-$c$ (a detailed derivation is provided in Supplementary Section 8). A key finding is that the linewidths of all three output spectra remain nearly identical to that of bare mode $c$ in the very weak-coupling regime where $g$ < 10 meV. As $g$ increases, this near equality persists over a broad weak-coupling region. These results demonstrate that the three output spectra are nearly indistinguishable under very weak coupling, indicating the emergence of a new degenerate mode whose linewidth closely matches that of mode $c$.

This convergence effect of the high-loss mode toward the low-loss mode remains valid even when the two modes are detuned. As shown in Fig. 3c, when the detuning ($\delta$ = $\omega_d - \omega_c$) is $-300$ meV, the linewidth and resonant frequency of the output spectrum from Ch-$d$ closely match those of the bare mode-$c$. Under detuned conditions, the linewidths and the frequency shifts of the Ch-$d$ and Ch-$c$ spectra relative to the bare mode-$c$ as a function of $g$ are plotted in Figs. 3d(i) and 3d(ii), demonstrating a robust convergence effect in the very weak-coupling regime. Furthermore, we plot these parameters over a detuning range of $\pm 2\Gamma_d$ at a fixed coupling strength of $g$ = 2 meV, as shown in Figs. 3e(i) and 3e(ii). Remarkably, even across a broad detuning range, the mode-$d$ is very well reshaped into the mode-$c$, indicating that the linewidth compression of the high-loss mode is highly robust and readily achievable in experiments. Notably, when the driving is switched to the high-loss Ch-$d$, the output spectra no longer exhibit linewidth compression. A detailed discussion is provided in Supplementary Section 9.

Our results demonstrate that when the narrow-linewidth mode is driven, it can reshape the broad-linewidth mode into a replica of itself in very weak-coupling regime. We believe that this convergence effect offers an effective strategy for suppressing the inherently large losses in localized plasmonic modes, and can be experimentally validated by constructing a non-Hermitian system in which a localized surface plasmonic mode is weakly coupled to a high-$Q$ dielectric resonant mode.

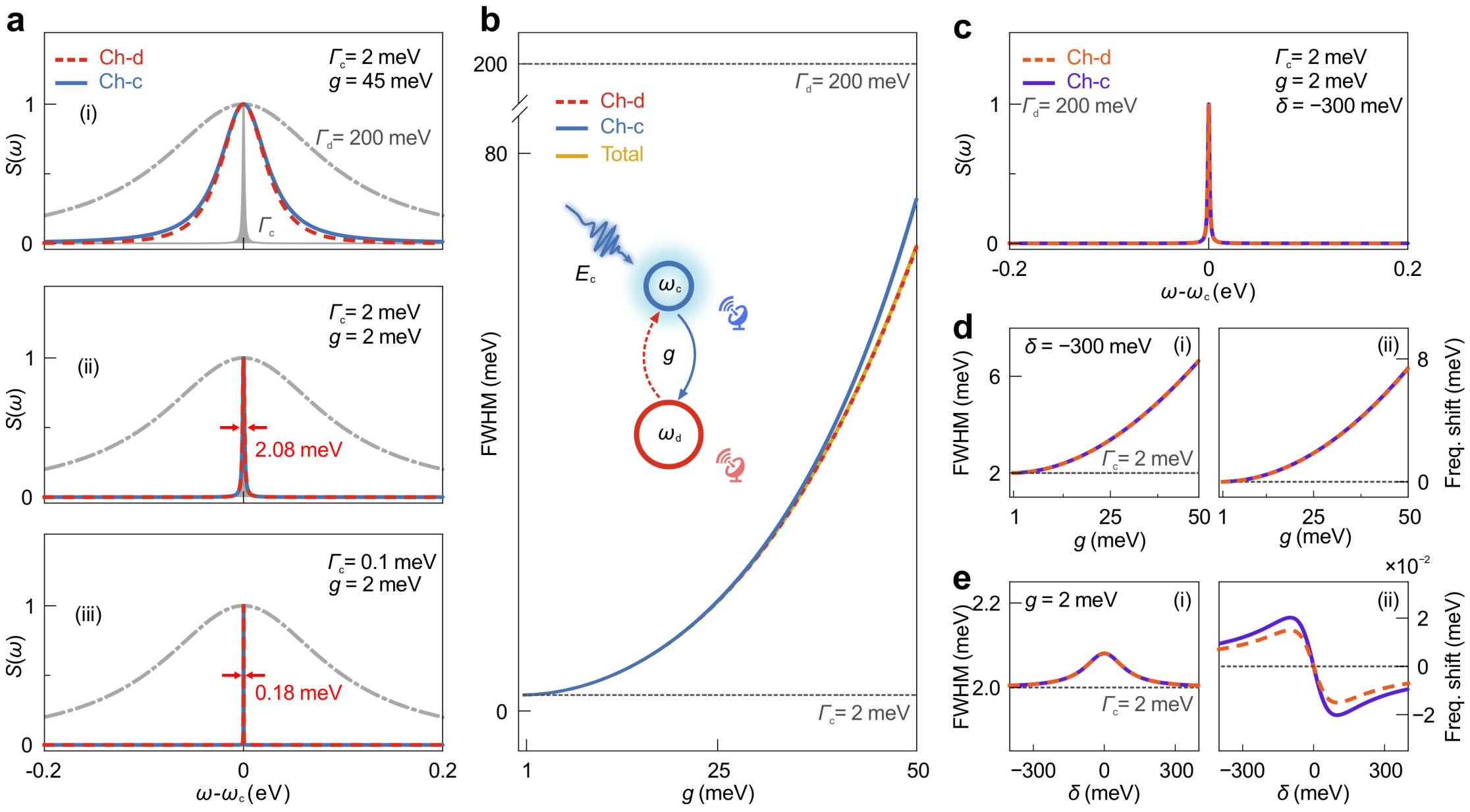

**Fig. 3 | Substantial suppression of modal losses. a**, Normalized output spectra from Ch-*d* and Ch-*c* under three different conditions. **b**, The FWHMs of the spectra from Ch-*d*, Ch-*c*, and total system as a function of $g$. **c**, Normalized output spectra from Ch-*d* and Ch-*c* at detuning $\delta = -300$ meV. **d, e,** The FWHMs and the frequency shifts of the Ch-*d* and Ch-*c* spectra relative to the bare mode-*c*, under various $g$ at $\delta = -300$ meV (d), and under various $\delta$ at $g = 2$ meV (e). In calculations, $E_d : E_c = 0{:}1$. The detuning case (panels c-e) is plotted using orange (Ch-*d*) and purple (Ch-*c*) curves.

## Counterintuitive disappearance of SRS

Conventionally, strong coupling leads to the formation of hybrid states in the coupled systems, characteristically manifested as the SRS. As illustrated in Figs. 4a and 4b, within the strong-coupling regime and provided that $g$ is sufficiently large, the spectra of both channels exhibit the clear SRS, irrespective of which mode is individually driven. However, under balanced driving configuration ($E_d$: $E_c$ = 1:1), the SRS in both channels vanishes, replaced by a single-peak lineshape that no longer reflects the intrinsic LS, as illustrated in Fig. 4c. This counterintuitive behavior also originates from the interference between the self-excitation and coupling-excitation spectral functions induced by weak external driving (Supplementary Section 10.1). This interference breaks the coupling symmetry, causing the originally symmetric Rabi-split spectra to first evolve into weak Fano-like lineshapes and eventually collapse into a single peak. For a fixed $\Gamma_d$, as $\Gamma_c$ gradually approaches $\Gamma_d$, the spectra progressively evolve from weak Fano lineshapes toward a standard Lorentzian lineshape, and the normalized spectra of the two channels become nearly identical. Moreover, this phenomenon is consistently observed in the total spectrum of the coupled system (Supplementary Section 10.2), indicating that the disappearance of the SRS is not a channel-specific artifact but a globally controllable effect induced by weak external driving.

Specifically, under the perfectly balanced conditions ($E_d : E_c$ = 1:1 and $\Gamma_d = \Gamma_c$), Fig. 4d shows that the output spectra of the two channels exhibit standard single-peaked Lorentzian lineshapes with identical central frequencies and linewidths, rendering the channels indistinguishable. In this scenario, as the coupling strength $g$ increases, the output spectra exhibit no SRS but instead undergo a giant continuous frequency shift with a fixed linewidth ($\Gamma_d$ or $\Gamma_c$) and without saturation (Supplementary Section 10.3). This large and controllable spectral shift, rather than the SRS, establishes a robust foundation for tunable light sources, high-precision quantum sensors, and frequency-agile photonic devices. Importantly, this tunability is achieved without linewidth broadening, overcoming the longstanding bottleneck in photonic devices, where frequency tuning typically degrades the quality factor.

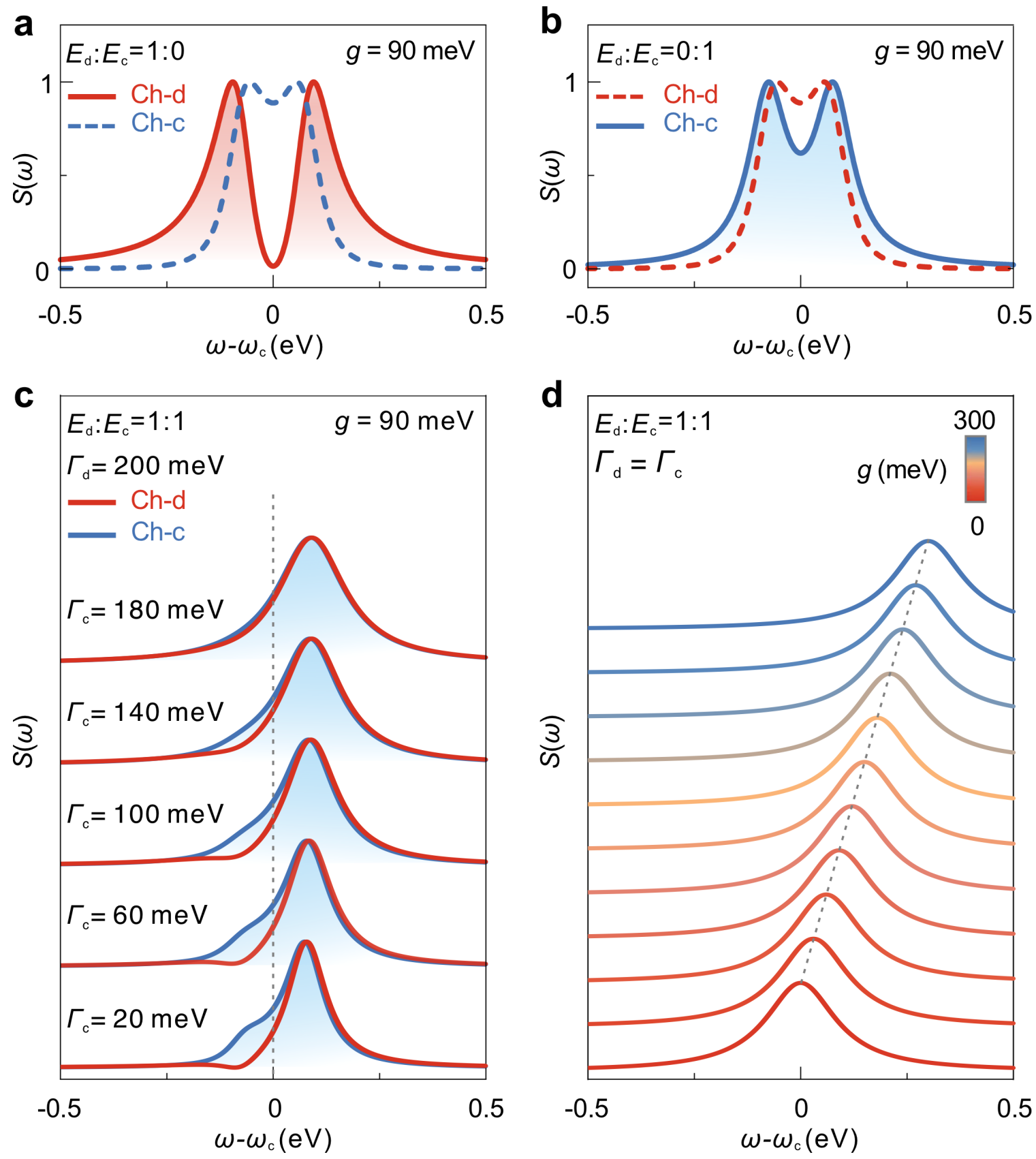

**Fig. 4 | Disappearance of SRS. a,b,** Normalized output spectra from Ch-*d* and Ch-*c*, corresponding to $E_d : E_c$ = 1:0 (**a**) and 0:1 (**b**), respectively, with $g$ = 90 meV. **c,** Normalized spectra from Ch-*d* and Ch-*c* with increasing $\Gamma_c$ for $E_d : E_c$ = 1:1 at a fixed $\Gamma_d$ = 200 meV. **d,** Spectral response for $\Gamma_d = \Gamma_c$ = 200 meV and $E_d : E_c$ = 1:1 at different $g$.

## Accurate characterization of LS

As discussed above and shown in Fig. 2, the output spectra generally fail to exhibit a fingerprint-like correspondence with the eigenlevels of coupled non-Hermitian systems. Particularly in the strong-coupling regime, the SRS is not equal to the LS, which has also been confirmed in our recent experiments[58]. It is important to establish a direct and straightforward method for experimentally measuring the intrinsic eigenlevels and the coupling strength $g$ of a coupled non-Hermitian system.

It is interesting to note from Eq. (4) that, when only one channel in the coupled system is driven (Fig. 5a), the output from the other channel exhibits an identical coupling-excitation spectral profile due to reciprocity, differing only in intensity. Moreover, its SRS at resonance in the strong-coupling regime is expressed as (Supplementary Section 11):

$$\Delta_{\mathrm{SRS},g} = 2\sqrt{g^2 - \frac{\Gamma_c^2+\Gamma_d^2}{8}}, \tag{5}$$

As shown in Fig. 5b, $\Delta_{\mathrm{SRS},g}$ is always smaller than $\Delta_{\mathrm{LS}}$ in Eq. (2). Importantly, we find that $\Delta_{\mathrm{LS}}$, $\Delta_{\mathrm{SRS},g}$, and the total average dissipation of the system satisfy the following Pythagorean relation at resonance (Fig. 5c):

$$\Delta_{\mathrm{LS}}^2 = \Delta_{\mathrm{SRS},g}^2 + \left(\frac{\Gamma_d+\Gamma_c}{2}\right)^2. \tag{6}$$

Therefore, the LS and $g$ can be directly and accurately determined from Eqs. (5) and (6) by experimentally measuring the SRS $\Delta_{\mathrm{SRS},g}$ in the coupling-excitation spectrum and the bare-mode linewidths $\Gamma_d$ and $\Gamma_c$. This offers a universal and practical approach to exactly measure the intrinsic eigenlevels of non-Hermitian coupled systems, establishing a new paradigm for eigenlevel characterization.

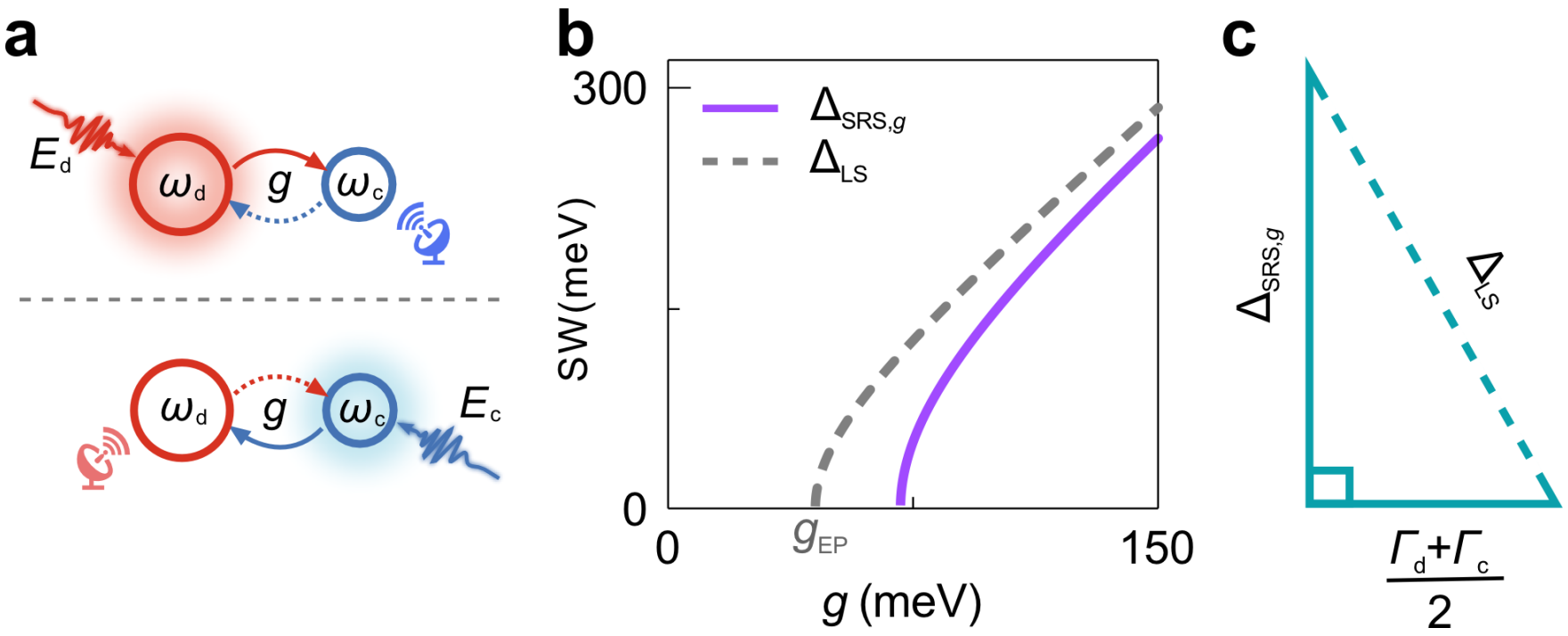

**Fig. 5 | Accurate characterization of LS. a,** Schematic of the coupled system under single-mode driving. **b,** Splitting width (SW) of the coupling-excitation spectra and the eigenlevels as a function of $g$. **c,** Pythagorean relation among $\Delta_{\mathrm{LS}}$, $\Delta_{\mathrm{SRS},g}$, and the average dissipation of the modes.

## Numerical simulations

To verify the validity of the excitation interference effect described by Eq. (4) in classical coupled systems, we designed a hybrid nanophotonic structure comprising a periodic silver nanopillar-array cavity (Ag NAC) coupled to a gold nanorod (Au NR), integrated on a multilayer substrate consisting of a $SiO_2$ spacer, an Ag film, and a semi-infinite $SiO_2$ substrate, as depicted in Fig. 6a. Full-wave simulations employing the finite-element method were performed (see Supplementary Section 12 for details). Under normal incidence illumination by an *x*-polarized plane wave, the Ag NAC functions as a lattice cavity supporting a high-*Q* surface lattice resonance mode, while the isolated Au NR exhibits a broadband localized surface plasmon resonance, as shown in Fig. 6b. The Au

NR resonance is slightly redshifted relative to the Ag NAC resonance to compensate for refractive-index variations, ensuring near-resonant coupling. Crucially, the Au NR experiences near-field modulation from the Ag NAC. The inset depicts the spatially inhomogeneous $|E_x|$ distribution of the resonant Ag NAC mode, which provides the physical mechanism for controlling the driving intensity ratio. As the Au NR position varies, its effective driving intensity correspondingly changes, enabling the continuous modulation of the driving intensity ratio through controlled nanorod displacement.

In our simulations, the Au NR was fixed at $\delta_y$ = 220 nm to mitigate the near-field scattering interference from the Ag nanopillar edges, while its lateral displacement $\delta_x$ relative to the nanopillar center was systematically varied to modulate the driving intensity ratio (gray arrow, Fig. 6c inset). Figure 6c presents the normalized absorption cross sections of the Au NR at different $\delta_x$ positions, corresponding to remarkably different driving intensity ratios and slightly different coupling strengths. These spectra successfully reproduce the theoretically predicted lineshape transitions. The excellent agreement between the simulation results and theoretical predictions conclusively confirms the applicability of our theoretical framework to both classical and quantum coupled systems.

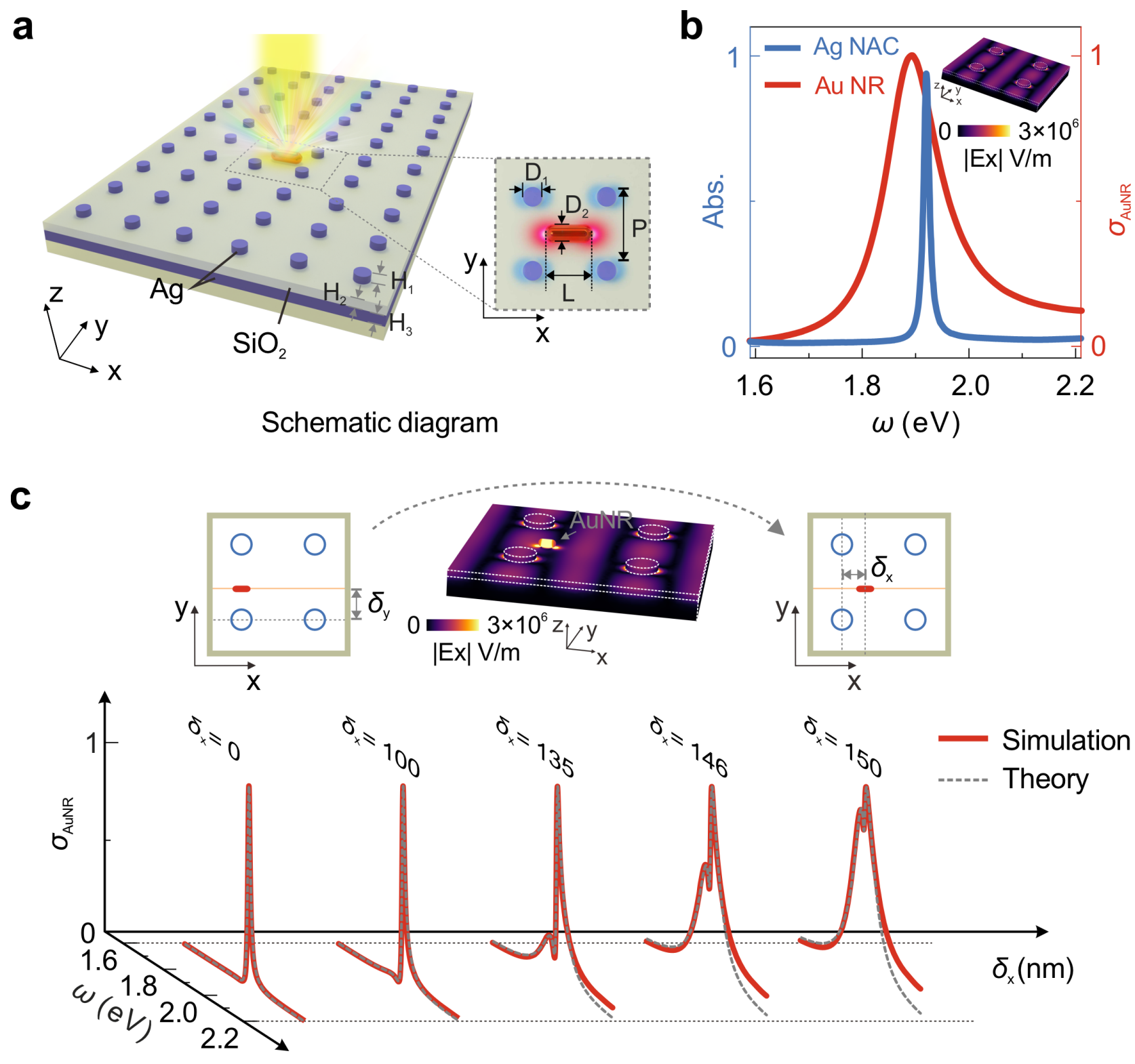

**Fig. 6 | Simulations of spectral modulation by driving-intensity ratio. a,** Schematic of

the hybrid system. **b,** Absorption spectrum of the Ag NAC ($\Gamma_c$ = 12.8 meV) and normalized absorption cross section ($\sigma_{\mathrm{AuNR}}$) of the uncoupled Au NR ($\Gamma_d$ = 118.5 meV). Inset: $|E_x|$ distribution of the Ag NAC at resonance (1.920 eV). **c,** Evolution of the normalized $\sigma_{\mathrm{AuNR}}$ at various lateral displacements $\delta_x$ (red curves) with the theoretical fits (gray curves). The variation in $\delta_x$ corresponds to tuning the driving intensity ratio ($E_d$: $E_c$ = 0.18:1, 0.41:1, 1:0.86, 1:0.32, 1:0.09) and the coupling strengths, which are $g$ = 9.1, 8.3, 7.8, 7.2, and 6.8 meV in the fitting. Center inset: $|E_x|$ distribution of the hybrid system at the absorption peak (1.922 eV) with $\delta_x$ = 100 nm. In simulations, $D_1$ = 160 nm, P = 600 nm, $H_1$ = $H_2$ = 30 nm, $H_3$ = 200 nm, $D_2$ = 40 nm and L = 92.5 nm.

## Conclusion

In summary, our work establishes a transformative framework for spectral manipulation in non-Hermitian coupled systems through weak external driving. Central to our discovery is interference-induced non-intrinsic spectral outputs due to dissipation: the weak external driving activates self-excitation and coupling-excitation spectral functions within linear response, their inteference outputs fundamentally decouple from intrinsic eigenstate mappings. This paradigm enables three pivotal capabilities: on-demand spectral engineering via driving ratio control; extreme dissipation suppression (> 1000-fold linewidth compression) through high-to-low-loss mode convergence effect; and giant spectral shift replacing original SRS under balanced driving. To resolve measurement distortion caused by spectral decoupling, we derive the resonant Pythagorean theorem connecting eigenlevel splitting, spectral Rabi splitting and total decay linewidth—estalishing the first direct protocol for experimentally measuring intrinsic parameters: level splitting and coupling strength. Validated across quantum-classical platforms, our work not only redefines the fundamental principles of measurement in non-Hermitian physics under weak external driving, but also finds immediate applications in plasmonic photonics, ultra-sensitive sensors, tunable nanoscale light sources, and room-temperature quantum devices.

## Data availability

Relevant data supporting the key findings of this study are available within the article and the Supplementary Information file. The original data generated during the current study are available from the corresponding authors upon request.

## Acknowledgment

This work was financially supported by the National Natural Science Foundations of China (Grant No. 12334017 and No. 12494602), the National Key R&D Program of China (Grant No. 2021YFA1400800), the Natural Science Foundation of Guangdong Province (Grant No. 2026A1515010709), and the Guangdong Provincial Quantum Science Strategic Initiative (Grant No. GDZX2406001).

## Author contributions

X.H.W. and W.L. presented the concept and conceived the research outline. H.F. and W.L. derived the formulae. H.F. and Y.F.X. conducted the numerical simulations. All the authors discussed the results and co-wrote the manuscript. X.H.W. and W.L. supervised the project.

## Competing interests

The authors declare no competing interests.

Supplementary information for:

# Manipulating non-intrinsic outputs of a non-Hermitian coupled system by weak external driving

He Feng[1#], Yi-Fan Xiang[1#], Wei Li[1*], and Xue-Hua Wang[1*]

[1]State Key Laboratory of Optoelectronic Materials and Technologies, School of Physics, Sun Yat-sen University, Guangzhou 510275, China

Corresponding author(s). E-mail(s): liwei373@mail.sysu.edu.cn; wangxueh@mail.sysu.edu.cn

[#]These authors contributed equally to this work

**Contents**

**Section 1. Fingerprint-like spectral correspondence of the probed system.**

In conventional spectral analysis of coupled systems, external driving is typically treated as weak excitation of a single mode or, more generally, assumed to play only a secondary role in the system dynamics. Within this framework, the external driving is regarded as a passive perturbative probe that excites the coupled system without altering its intrinsic spectral line-shape. Consequently, the observed spectral response is determined solely by the system's intrinsic properties, such as dissipation, coupling and detuning. This establishes a fingerprint-like correspondence between the measured spectrum and the underlying eigen-structure, enabling intrinsic system parameters to be inferred directly from the spectral response. Spectroscopy therefore serves as a powerful diagnostic tool for characterizing coupled systems.

For instance, in plasmon-exciton hybrid systems, the absorption spectra from plasmonic channel (*d*-channel) can be obtained using the method of Zubarev's Green Functions when only the *d*-channel is excited [1, 2]:

$$\sigma_{\mathrm{abs},d} \propto -\mathrm{Im}\left\{\frac{\omega-\omega_c+i\dfrac{\Gamma_c}{2}}{\left(\omega-\omega_d+i\dfrac{\Gamma_d}{2}\right)\left(\omega-\omega_c+i\dfrac{\Gamma_c}{2}\right)-g^2}\right\}. \tag{S-1}$$

Where $\omega_i$ ($i$ = $d$, $c$) represent the resonant frequencies of the modes in the coupled system, $\Gamma_i$ denote the dissipative linewidths of the modes, and $g$ is the coupling strength between the modes. Conversely, when only the excitonic channel (*c*-channel) is excited, the emission spectra collected from the two output channels can be derived from the corresponding two-time correlation functions [3]:

$$S_{\mathrm{PL},c} \propto \left|\frac{\dfrac{\Gamma_d}{2}-i(\omega-\omega_d)}{\left[\dfrac{\Gamma_d+\Gamma_c}{4}+i\left(\dfrac{\omega_c-\omega_d}{2}\right)-i(\omega-\omega_d)\right]^2+g_{\mathrm{eff}}^2}\right|^2, \tag{S-2}$$

$$S_{\mathrm{PL},d} \propto \left|\frac{-ig}{\left[\dfrac{\Gamma_d+\Gamma_c}{4}+i\left(\dfrac{\omega_c-\omega_d}{2}\right)-i(\omega-\omega_d)\right]^2+g_{\mathrm{eff}}^2}\right|^2, \tag{S-3}$$

where $g_{\mathrm{eff}}=\sqrt{g^2+(\omega_d/2-\omega_c/2)^2-(\Gamma_d/4-\Gamma_c/4)^2}$.

In both absorption and emission spectra, the spectral response is governed entirely by these intrinsic system parameters, whereas the external driving contributes only an overall intensity scaling and does not participate in shaping the spectral line-shape. Thus, the spectral line-shape remains a fingerprint-like signature of the underlying eigen-structure. However, this conventional picture breaks down when both modes of a non-Hermitian coupled system are driven simultaneously, where external driving opens interference excitation pathways absent under single-mode driving. The corresponding theoretical framework is developed in the main text.

**Section 2. Analytical derivation of the output spectra under external driving.**

Under weak external driving, we analytically investigate the spectral responses of coupled bosonic systems, coupled fermion-boson systems, and single-mode driving.

**2.1 Spectra of coupled bosonic system.**

We first consider a generic coupled bosonic system comprising two bosonic modes, *d* and *c,* with resonance frequencies $\omega_d$ and $\omega_c$, respectively, and coupling strength $g$. Under weak external driving, the system is described by a total Hamiltonian that can be decomposed into an effective system Hamiltonian and a driving Hamiltonian,

$$H = H_{\text{sys}} + H_{\text{dri}}. \tag{S-4}$$

Where

$$H_{\text{sys}} = \left(\omega_d - i\frac{\Gamma_d}{2}\right)d^+ d + \left(\omega_c - i\frac{\Gamma_c}{2}\right)c^+ c + g(d^+ c + c^+ d), \tag{S-5}$$

$$H_{\text{dri}} = E_d\left(d^+ + d\right) + E_c\left(c^+ + c\right). \tag{S-6}$$

The time evolution of the operator *d* and *c* is governed by:

$$\frac{d}{dt}d = -i[d, H] = -i[d, H_{\text{sys}} + H_{\text{dri}}], \tag{S-7}$$

$$\frac{d}{dt}c = -i[c, H] = -i[c, H_{\text{sys}} + H_{\text{dri}}]. \tag{S-8}$$

Then, we obtain the full Heisenberg equation of motion for the operator $d$ and $c$

$$\frac{d}{dt}d = -i\left(\omega_d - i\frac{\Gamma_d}{2}\right)d - igc - iE_d, \tag{S-9}$$

$$\frac{d}{dt}c = -i\left(\omega_c - i\frac{\Gamma_c}{2}\right)c - igd - iE_c. \tag{S-10}$$

Applying the Laplace transform to both sides of the equations of motion, we obtain

$$sd(s) = -i\left(\omega_d - i\frac{\Gamma_d}{2}\right)d(s) - igc(s) - iE_d(s), \tag{S-11}$$

$$sc(s) = -i\left(\omega_c - i\frac{\Gamma_c}{2}\right)c(s) - igd(s) - iE_c(s). \tag{S-12}$$

Transforming to the frequency domain by setting $s = -i\omega$ yields

$$-i\omega d = -i\left(\omega_d - i\frac{\Gamma_d}{2}\right)d - igc - iE_d, \tag{S-13}$$

$$-i\omega c = -i\left(\omega_c - i\frac{\Gamma_c}{2}\right)c - igd - iE_c. \tag{S-14}$$

Therefore

$$d = E_d\psi_d + E_c\psi_g, \tag{S-15}$$

$$c = E_c\psi_c + E_d\psi_g. \tag{S-16}$$

Where $\psi_{d(c)} = (\omega - \omega_{c(d)} + i\Gamma_{c(d)}/2)/\chi$ and $\psi_g = g/\chi$ denote the channels' self-excitation and coupling-excitation spectral functions, respectively; $\chi = (\omega - \omega_d + i\Gamma_d/2)(\omega - \omega_c + i\Gamma_c/2) - g^2 = (\omega - \omega_+)(\omega - \omega_-)$.

The spectral intensity of the coupled system from Ch-*d* and Ch-*c* is

$$S_d \propto \langle d^+ d\rangle = |E_d\psi_d + E_c\psi_g|^2, \tag{S-17}$$

$$S_c \propto \langle c^+ c\rangle = |E_c\psi_c + E_d\psi_g|^2. \tag{S-18}$$

It follows that, when both modes are driven simultaneously, the observed spectral response arises from the interference between the self-excitation and coupling-excitation spectral functions. Taking into account the spectral weight associated with each detection channel, the total spectrum of the coupled system can be expressed as

$$S_{\text{total}} \propto A_d S_d + A_c S_c. \tag{S-19}$$

In general, the coefficients $A_d = \Gamma_d/2$ and $A_c = \Gamma_c/2$ [4].

## 2.2 Spectra of coupled fermion-boson system.

Next, we consider a coupled system involving a fermionic mode. Assuming that mode *c* is fermionic, analogous to S2.1, the governing equation for the bosonic mode *d* is given by

$$\frac{d}{dt}d = -i\left(\omega_d - i\frac{\Gamma_d}{2}\right)d - igc - iE_d, \tag{S-20}$$

and the Heisenberg equation of motion for the operator $c$ is

$$\frac{d}{dt}c=-i\left(\omega_c-i\frac{\Gamma_c}{2}\right)c-ig\left(1-2c^+c\right)d-i\left(1-2c^+c\right)E_c. \qquad \text{(S-21)}$$

Further manipulation of this equation would generate an infinite hierarchy of equations of motion. However, we terminate the sequence at this stage by applying an approximation of $\langle c^+c\rangle = n_c$. The emergence of this specific term stems from the inherent fermionic nature of the quantum emitter. As a result,

$$\frac{d}{dt}c=-i\left(\omega_c-i\frac{\Gamma_c}{2}\right)c-ig\left(1-2n_c\right)d-i\left(1-2n_c\right)E_c. \qquad \text{(S-22)}$$

Applying the Laplace transform to both sides of the equations of motion Eqs. (S-20) and (S-22), we obtain

$$sd\left(s\right)=-i\left(\omega_d-i\frac{\Gamma_d}{2}\right)d(s)-igc(s)-iE_d(s), \qquad \text{(S-23)}$$

$$sc(s)=-i\left(\omega_c-i\frac{\Gamma_c}{2}\right)c(s)-i\left(1-2n_c\right)gd(s)-i\left(1-2n_c\right)E_c(s). \qquad \text{(S-24)}$$

Transforming to the frequency domain by setting $s = -i\omega$ yields

$$-i\omega d=-i\left(\omega_d-i\frac{\Gamma_d}{2}\right)d-igc-iE_d, \qquad \text{(S-25)}$$

$$-i\omega c=-i\left(\omega_c-i\frac{\Gamma_c}{2}\right)c-i\left(1-2n_c\right)gd-i\left(1-2n_c\right)E_c. \qquad \text{(S-26)}$$

Then the operators $d$ and $c$ can be expressed by

$$d=E_d\psi'_d+\left(1-2n_c\right)E_c\psi'_g, \qquad \text{(S-27)}$$

$$c=\left(1-2n_c\right)\left(E_c\psi'_c+E_d\psi'_g\right). \qquad \text{(S-28)}$$

Where $\psi'_{d(c)} = (\omega - \omega_{c(d)} + i\Gamma_{c(d)}/2)/\chi'$ and $\psi'_g = g/\chi'$denote the channels' self-excitation and coupling-excitation spectral functions, respectively; $\chi' = (\omega - \omega_d + i\Gamma_d/2)(\omega - \omega_c + i\Gamma_c/2) - g^2(1-2n_c)$. Accordingly, the output spectral intensity of the coupled system from Ch-*d* and Ch-*c* is

$$S_d \propto \langle d^+d\rangle=\left|E_d\psi'_d+\left(1-2n_c\right)E_c\psi'_g\right|^2, \qquad \text{(S-29)}$$

$$S_c \propto \langle c^+c\rangle=\left(1-2n_c\right)^2\left|\left(E_c\psi'_c+E_d\psi'_g\right)\right|^2. \qquad \text{(S-30)}$$

In the absence of nonlinearity, the system is assumed to be initially prepared in its ground state, i.e., $n_c = 0$. Under these conditions, the spectrum is identical to that of the bosonic system.

## 2.3 Spectral response under weak single-mode driving.

We next consider the spectral response under weak single-mode driving. When the driving term acts only on mode $d$ ($E_d : E_c = 1 : 0$), as shown in Fig. S1(a), the out spectra obtained from Ch-$d$ (self-excitation) and Ch-$c$ (coupling-excitation) are given by

$$S_d^{\mathrm{dri},d} \propto \langle d^+ d \rangle = | E_d \psi_d |^2, \tag{S-31}$$

$$S_c^{\mathrm{dri},d} \propto \langle c^+ c \rangle = | E_d \psi_g |^2. \tag{S-32}$$

As the driving term acts only on mode $c$ ($E_d : E_c = 0 : 1$), as shown in Fig. S1(b), the out spectra of the coupled system from Ch-$c$ (self-excitation) and Ch-$d$ (coupling-excitation) are given by

$$S_c^{\mathrm{dri},c} \propto \langle c^+ c \rangle = | E_c \psi_c |^2, \tag{S-33}$$

$$S_d^{\mathrm{dri},c} \propto \langle d^+ d \rangle = | E_c \psi_g |^2. \tag{S-34}$$

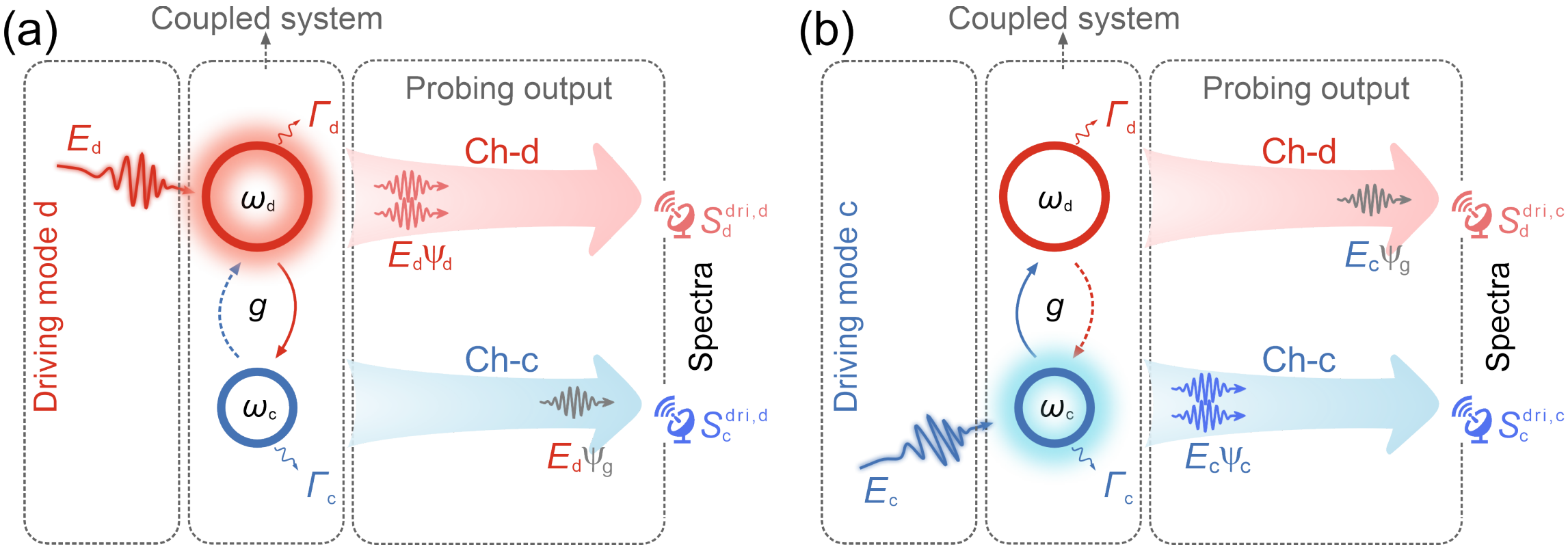

FIG. S1. Output spectra of the coupled system under single-mode driving: only mode $d$ is driven in (a), and only mode $c$ is driven in (b).

For comparison, the Hamiltonians of the bare modes $d$ and $c$ under driving are given by the following two expressions:

$$H_d = \left( \omega_d - i \frac{\Gamma_d}{2} \right) d^+ d + E_d \left( d^+ + d \right), \tag{S-35}$$

$$H_c = \left( \omega_c - i \frac{\Gamma_c}{2} \right) c^+ c + E_c \left( c^+ + c \right). \tag{S-36}$$

Similarly, by solving the Heisenberg equations of motion and applying the Laplace transform, the frequency-domain spectra of the bare modes $d$ and $c$ can be obtained as follows:

$$S_{d0} \propto \langle d^{+}d \rangle = \left| \frac{E_d}{\omega - \omega_d + i\frac{\Gamma_d}{2}} \right|^2 , \tag{S-37}$$

$$S_{c0} \propto \langle c^{+}c \rangle = \left| \frac{E_c}{\omega - \omega_c + i\frac{\Gamma_c}{2}} \right|^2 . \tag{S-38}$$

Importantly, this behavior is fully consistent with our analytical results: in the single-mode driving regime, the coupled-system expressions in Eqs. (S-17) and (S-18) reduce exactly to the $g = 0$ limit, recovering the purely Lorentzian response of uncoupled modes. This agreement confirms the validity and broad applicability of the present theoretical framework.

**Section 3. Spectral symmetry in conventional theory.**

In the conventional theoretical treatment of spectral responses in non-Hermitian coupled systems[3, 5] — equivalently, in the single-mode driving limit of our framework—the output spectra of the two detection channels at resonance ($\omega_d = \omega_c = \omega_0$) are governed by the self-excitation spectrum and the coupled-excitation spectrum, respectively:

$$S_d^{\mathrm{dri},d} \propto \frac{(\omega-\omega_0)^2 + \frac{\Gamma_c^2}{4}}{\left[(\omega-\omega_0)^2 - \frac{\Gamma_d \Gamma_c}{4} - g^2\right]^2 + \left[(\omega-\omega_0)\frac{\Gamma_c+\Gamma_d}{2}\right]^2} , \tag{S-39a}$$

$$S_c^{\mathrm{dri},d} \propto \frac{g^2}{\left[(\omega-\omega_0)^2 - \frac{\Gamma_d \Gamma_c}{4} - g^2\right]^2 + \left[(\omega-\omega_0)\frac{\Gamma_c+\Gamma_d}{2}\right]^2} . \tag{S-39b}$$

$$S_c^{\mathrm{dri},c} \propto \frac{(\omega-\omega_0)^2 + \frac{\Gamma_d^2}{4}}{\left[(\omega-\omega_0)^2 - \frac{\Gamma_d \Gamma_c}{4} - g^2\right]^2 + \left[(\omega-\omega_0)\frac{\Gamma_c+\Gamma_d}{2}\right]^2} , \tag{S-40a}$$

$$S_d^{\mathrm{dri},c} \propto \frac{g^2}{\left[(\omega-\omega_0)^2 - \frac{\Gamma_d \Gamma_c}{4} - g^2\right]^2 + \left[(\omega-\omega_0)\frac{\Gamma_c+\Gamma_d}{2}\right]^2} . \tag{S-40b}$$

In contrast to Eqs. (S-17) and (S-18), the spectral output in this scenario lacks interference between self-excitation and coupled-excitation spectral functions. As a result, the spectrum exhibits a symmetric line-shape centered at the resonance frequency $\omega_0$.

Figure S2 plots the spectra calculated using Eqs. (S-39) and (S-40) when the high-dissipation and the low-dissipation modes are individually driven, respectively, under both weak- and strong-coupling conditions. In all cases, regardless of whether the spectra exhibit a single peak or splitting, the line-shapes remain symmetric. In addition, the total absorption spectra calculated using Eq. (S-1) under single-mode driving also remains lineshape-symmetric, as demonstrated in our previous work [2].

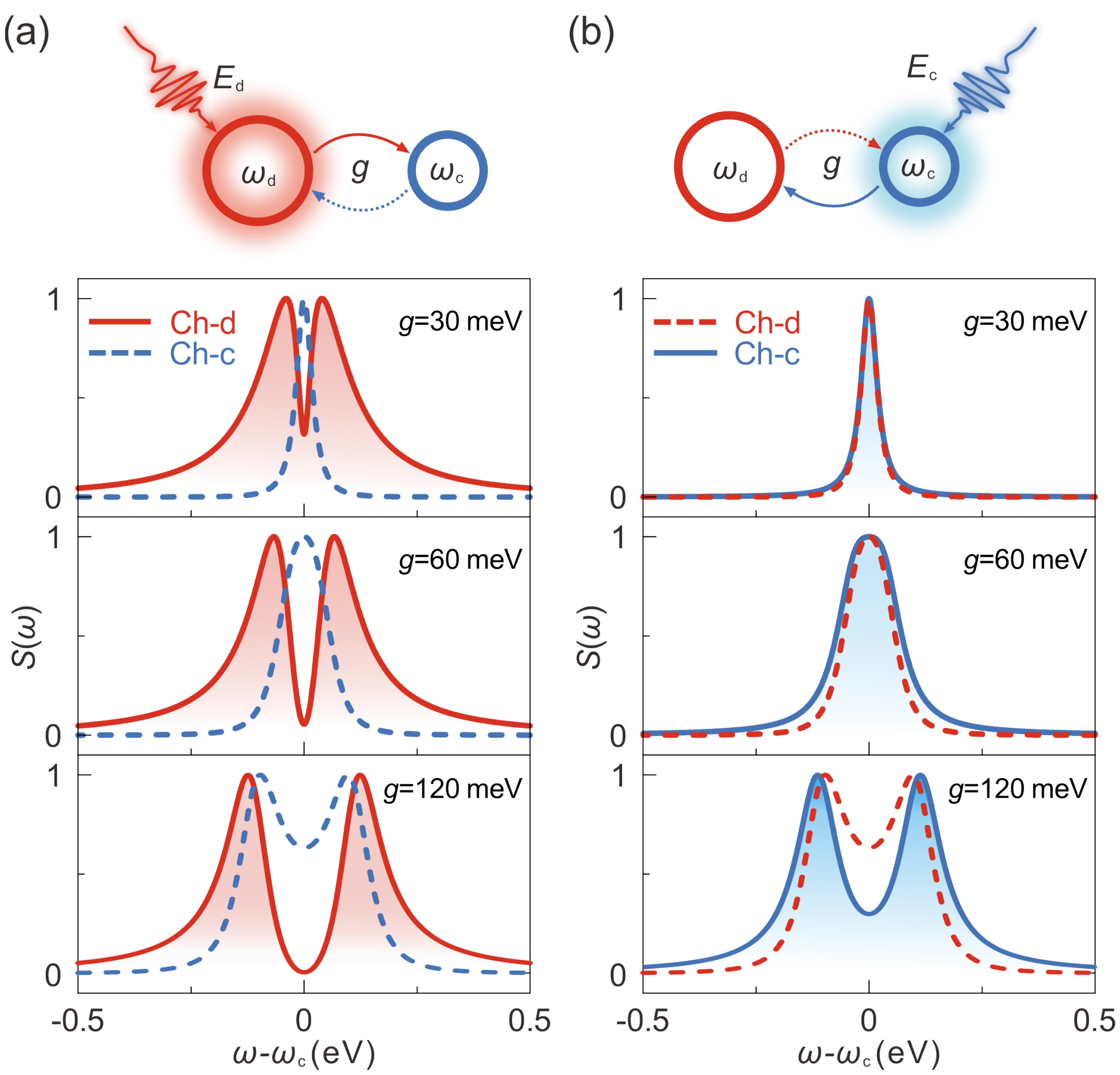

FIG S2. Normalized spectra calculated using Eqs. (S-39) and (S-40) for coupling strengths $g$= 30, 60, and 120 meV, when high-dissipation mode (a) and low-dissipation mode (b) are driven, respectively. In calculations, $\Gamma_d$ = 200 meV and $\Gamma_c$ = 20 meV.

**Section 4. Modulation of the system's total spectra.**

In this section, we investigate how the driving intensity ratio modulates the total spectral response under resonant conditions. For fixed coefficients $A_d = \Gamma_d/2$ and $A_c = \Gamma_c/2$, the total spectral response of the coupled system can be written as

$$\begin{aligned} S_{\text{total}} &\propto \Gamma_d S_d/2 + \Gamma_c S_c/2 \\ &= \frac{\Gamma_d}{2}\left|\frac{\left(\omega-\omega_c+i\frac{\Gamma_c}{2}\right)E_d+gE_c}{\chi}\right|^2+\frac{\Gamma_c}{2}\left|\frac{\left(\omega-\omega_d+i\frac{\Gamma_d}{2}\right)E_c+gE_d}{\chi}\right|^2. \end{aligned} \quad \text{(S-41)}$$

As evident from the above expression, the relative intensities of the external driving fields govern both the intra-channel interference and the inter-channel superposition, thereby fundamentally modifying the total spectral response. Consequently, tuning the driving intensity ratio enables substantial reshaping of the spectral profile, providing a powerful means for actively engineering the total output optical field.

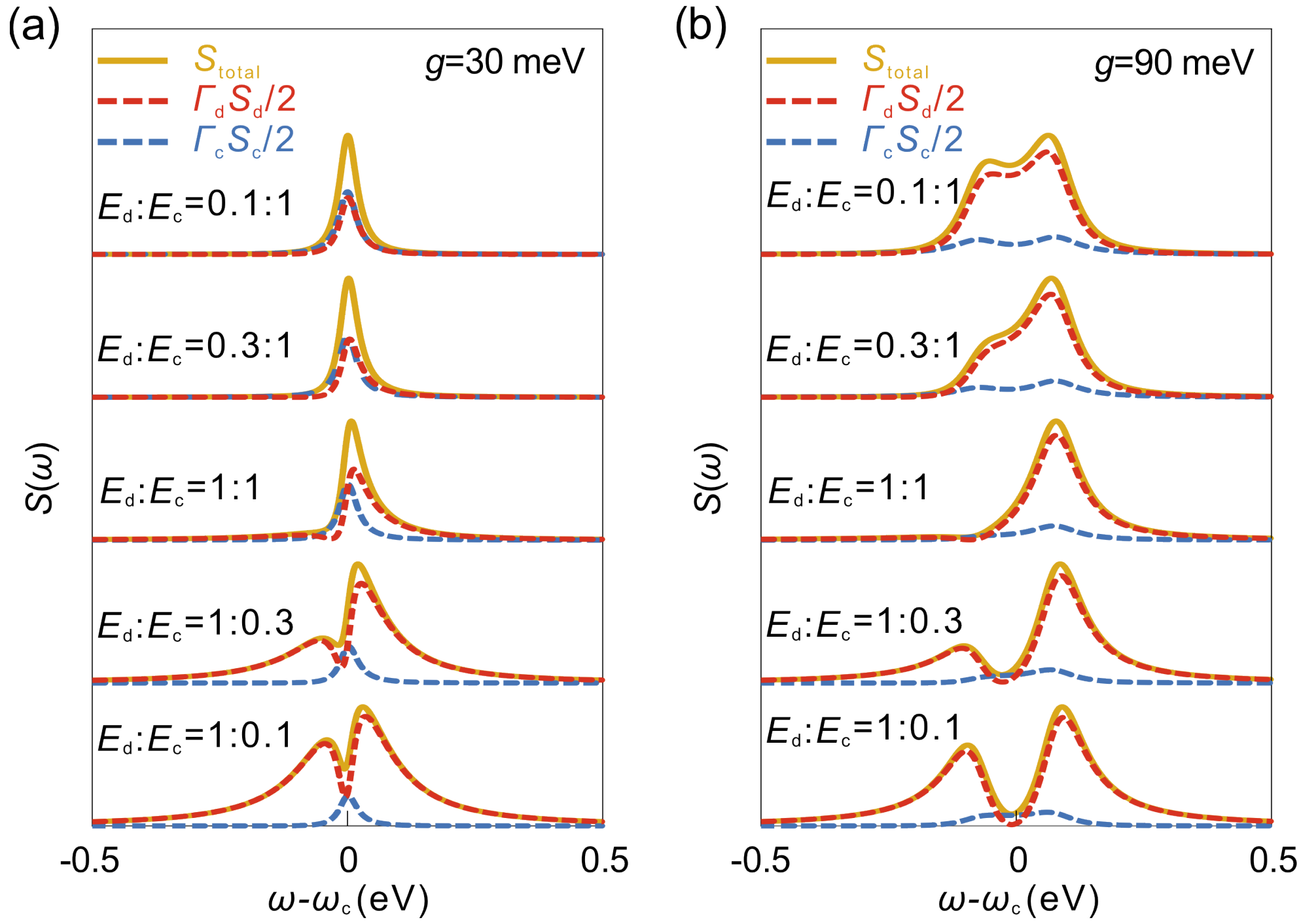

FIG. S3. The normalized total spectra of the coupled system (solid curves) under different $E_d$: $E_c$ with $g = 30$ (a), and 90 (b) meV. The dashed curves represent the results of the spectra from each channel normalized to the maximum value of the total spectrum after accounting for the coefficient $A_d = \Gamma_d/2$ and $A_c = \Gamma_c/2$. In calculations, $\Gamma_d$ = 200 meV and $\Gamma_c$ = 20 meV.

Figure S3 illustrates the resulting spectral evolution as the driving intensity ratio is varied for coupling strengths of $g$ = 30 and 90 meV, respectively. In all cases, tuning the driving ratio induces pronounced modifications of the total spectral response, highlighting the crucial role of relative driving strength in spectral control. In the weak-coupling regime, this tuning leads to diverse line shapes, including narrowband, Fano, and various asymmetric splitting features [Fig. S3(a)]. In the strong-coupling regime, the driving intensity ratio provides an additional degree of freedom to tailor the output spectra, enabling flexible control over the linewidth and splitting depth, the transition between single-peak and split states, and the emergence of asymmetric splitting patterns [Fig. S3(b)].

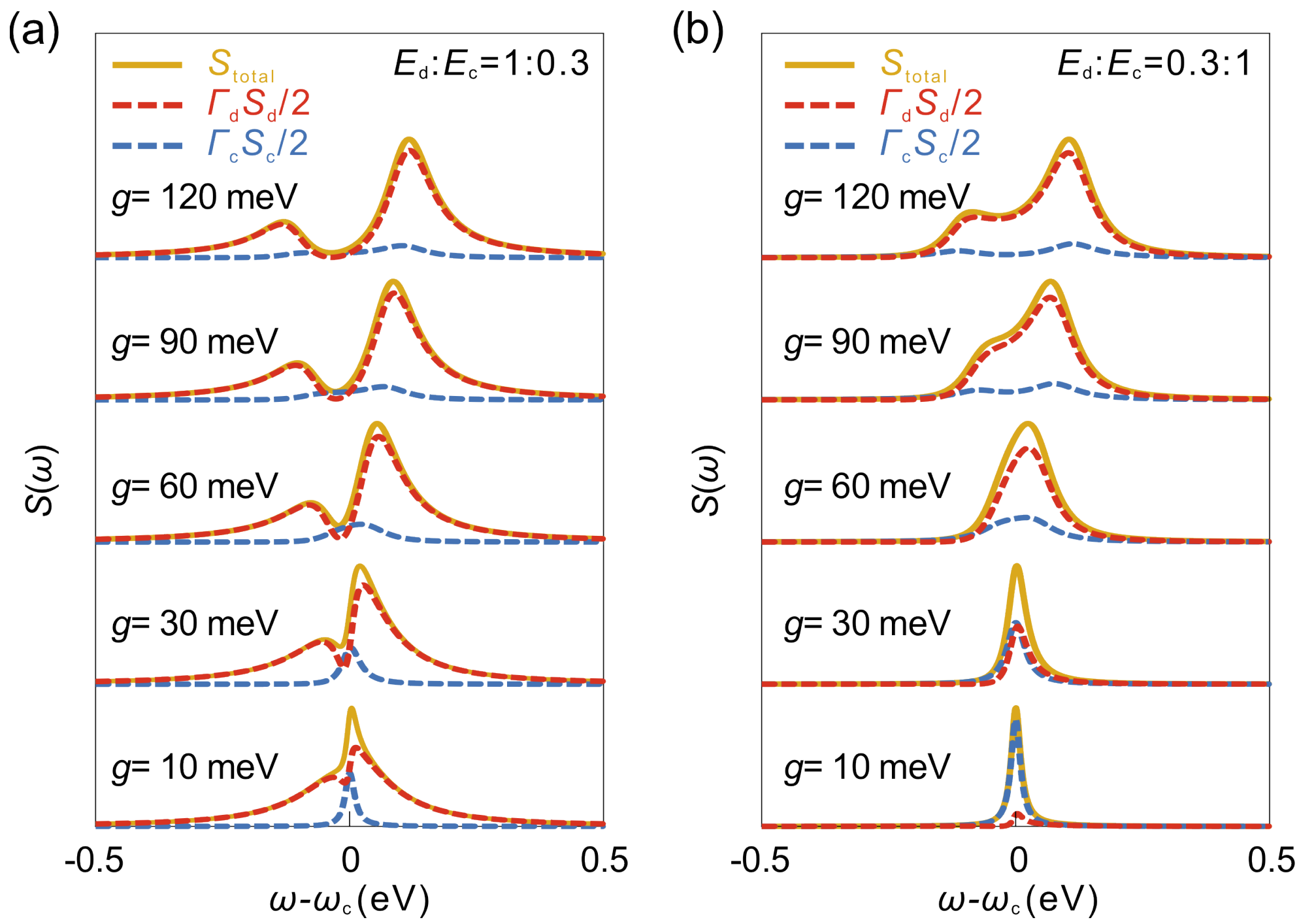

FIG. S4. The total normalized spectra of the coupled system (solid curves) under $E_d$: $E_c$ = 1:0.3 (a), and 0.3:1 (b) with different $g$. The dashed curves represent the results of the spectra from each channel normalized to the maximum value of the total spectrum after accounting for the coefficient $A_d = \Gamma_d/2$ and $A_c = \Gamma_c/2$. In calculations, $\Gamma_d$ = 200 meV and $\Gamma_c$ = 20 meV.

By fixing the driving intensity ratio, we explore the dependence of the total spectral response on the coupling strength. As shown in Fig. S4, increasing the coupling strength leads to a progressive convergence of the total spectrum toward the broadband mode channel, reflecting the increasing dominance of this channel in the overall spectral formation. Importantly, at different driving intensity ratios, the evolution of the total spectral response of the coupled system across the weak-to-strong coupling regime is significantly modulated. When the driving is predominantly applied to the high-loss mode $d$, increasing the coupling strength drives a transition of the total output spectral line-shape from a broad single-peak profile to an asymmetric splitting [Fig. S4(a)]. In contrast, when the driving is predominantly applied to the low-loss mode $c$, increasing the coupling strength drives a transition of the total output spectral line-shape from narrowband to asymmetric splitting [Fig. S4(b)].

Figures S3 and S4 reveal that the driving intensity ratio plays a central role in governing and modulating the total spectral response of the system. Notably, by optimizing the experimental detection scheme to precisely control the weighting of channel-resolved spectra, a richer set of superposition configurations can be accessed. This provides an additional degree of freedom for tailoring the total optical field output of the coupled system.

**Section 5. Spectral evolution versus coupling strength $g$ at different $E_d : E_c$.**

To further elucidate the interplay between coupling strength and driving intensity ratio, we investigate the evolution of the output spectra with increasing coupling strength at fixed driving intensity ratios. As shown in Fig. S5(a), for the Ch-$d$ output, when high-loss mode $d$ is predominantly excited, the spectra exhibit an asymmetric spectral splitting, with the splitting progressively widening as the coupling strength increases. In contrast, under driving predominantly applied to the low-loss mode $c$, the spectra evolve from a narrowed Fano resonance to asymmetric splitting. Notably, the initially narrowed Fano resonance demonstrates remarkably robust against variations in coupling strength $g$. Figure S5(b) illustrates the spectral evolution of Ch-$c$ versus $g$ at different $E_d : E_c$. Under high-loss mode-$d$-dominated excitation, the spectra initially exhibit pronounced broadening followed by a bifurcation into two asymmetric peaks. Dominating the driving on Ch-$c$ will accelerate this process. In addition, it can be observed from Fig. S5 that when the driving intensities of both modes are comparable, the spectral asymmetry becomes extremely high due to interference effects, which is even more pronounced in the strong coupling regime. Consequently, under different driving intensity ratios, tuning the coupling strength enables rich spectral reconfiguration, offering a versatile platform for dynamic optical-field tailoring.

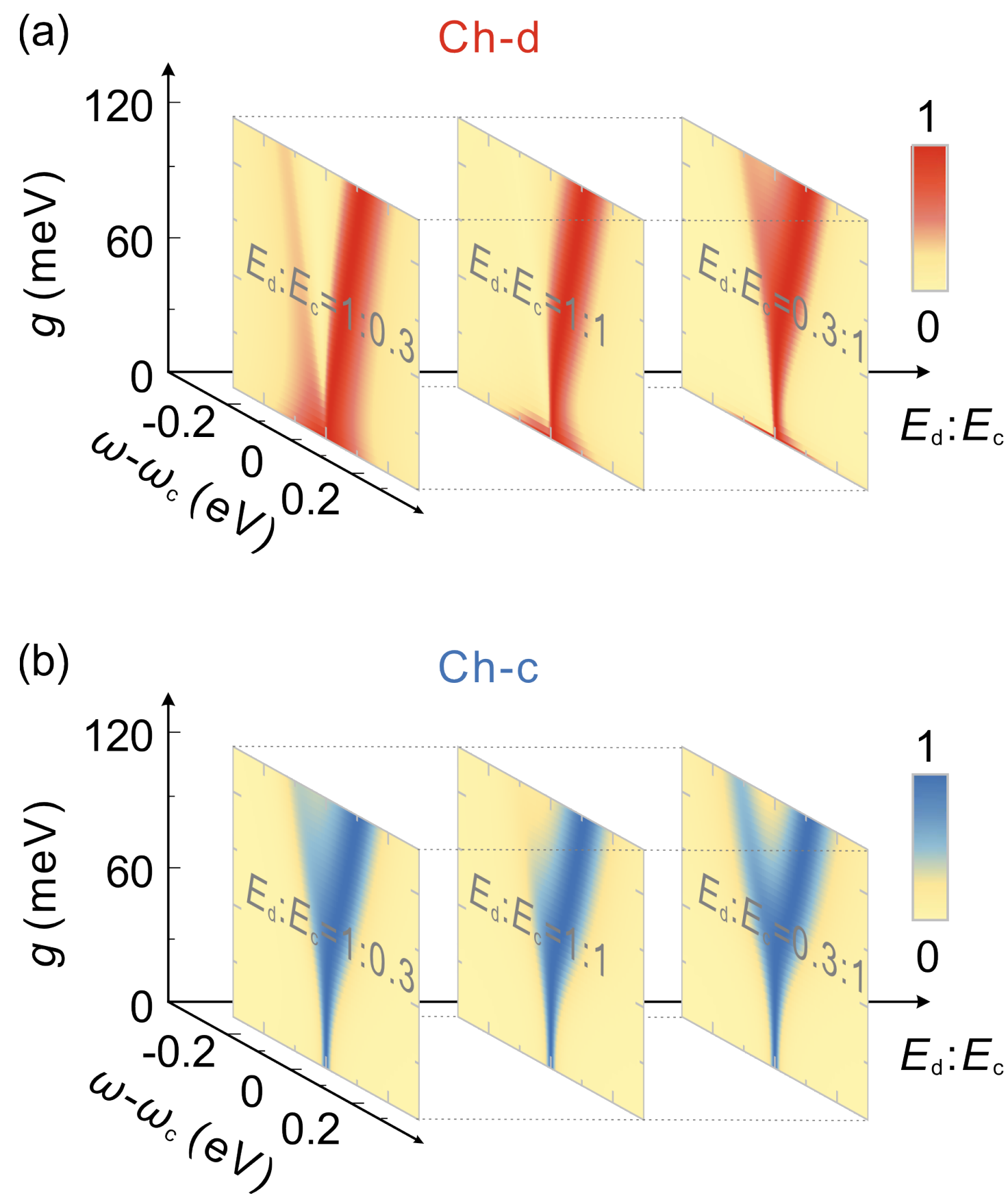

FIG. S5. Spectra of the Ch-$d$ (a) and Ch-$c$ (b) versus $g$ for $E_d$:$E_c$ = 1:0.3, 1:1, and 0.3:1. In calculations, $\Gamma_d$ = 200 meV and $\Gamma_c$ = 20 meV.

**Section 6. Previously reported theories as limiting cases of our unified framework.**

To highlight the generality of the present framework, we demonstrate that previously reported theories emerge naturally as limiting cases of our formalism. By imposing the corresponding single-channel driving conditions on the general expressions, the spectral properties reported in earlier studies can be recovered straightforwardly, thereby establishing a unified description of both single- and dual-channel excitation schemes.

According to Section 2.3, when only a single mode of the coupled system is externally driven (Fig. S1), our theory reduces to Eqs. (S-31)–(S-34). Notably, Eqs. (S-33) and (S-34) recover the same analytical form as the side-emission and forward-emission spectra reported in Ref. [3], while Eq. (S-33) additionally recovers the spontaneous-emission spectrum derived in Ref. [6]. Thus, the analytical expressions for emission spectra reported in previous studies are naturally recovered from the present framework under the corresponding single-channel driving conditions, confirming that these earlier theories are limiting cases of our unified formalism. This framework can also be applied to analyze the absorption and radiative output power of microcavity-engineered plasmonic resonances [5].

Furthermore, by substituting the coefficients $A_d = \Gamma_d/2$ and $A_c = \Gamma_c/2$ into the total spectrum formula $S_{\text{total}} = A_d S_d + A_c S_c$ [i.e., Eq. (S-19)], the total spectrum in the single-channel driving limit can be obtained as

$$S_{\text{total}}^{\text{dri},d} \propto \frac{\Gamma_d}{2} S_d^{\text{dri},d} + \frac{\Gamma_c}{2} S_c^{\text{dri},d}, \tag{S-42a}$$

$$S_{\text{total}}^{\text{dri},c} \propto \frac{\Gamma_d}{2} S_d^{\text{dri},c} + \frac{\Gamma_c}{2} S_c^{\text{dri},c}. \tag{S-42b}$$

Expanding the formula yields

$$S_{\text{total}}^{\text{dri},d} \propto |E_d|^2 \frac{\frac{\Gamma_d}{2}\left[(\omega-\omega_c)^2 + \frac{\Gamma_c^2}{4}\right] + \frac{\Gamma_c}{2} g^2}{\left[(\omega-\omega_d)(\omega-\omega_c) - \frac{\Gamma_d \Gamma_c}{4} - g^2\right]^2 + \left[(\omega-\omega_d)\frac{\Gamma_c}{2} + (\omega-\omega_c)\frac{\Gamma_d}{2}\right]^2}, \tag{S-43a}$$

$$S_{\text{total}}^{\text{dri},c} \propto |E_c|^2 \frac{\frac{\Gamma_c}{2}\left[(\omega-\omega_d)^2 + \frac{\Gamma_d^2}{4}\right] + \frac{\Gamma_d}{2} g^2}{\left[(\omega-\omega_d)(\omega-\omega_c) - \frac{\Gamma_d \Gamma_c}{4} - g^2\right]^2 + \left[(\omega-\omega_d)\frac{\Gamma_c}{2} + (\omega-\omega_c)\frac{\Gamma_d}{2}\right]^2}. \tag{S-43b}$$

Next, we consider the conventional absorption spectrum derived using the Zubarev's Green's function formalism. Expanding Eq. (S-1) yields [1, 2] :

$$\sigma_{\text{abs},d} \propto \frac{\frac{\Gamma_d}{2}\left[(\omega-\omega_c)^2 + \frac{\Gamma_c^2}{4}\right] + \frac{\Gamma_c}{2} g^2}{\left[(\omega-\omega_d)(\omega-\omega_c) - \frac{\Gamma_d \Gamma_c}{4} - g^2\right]^2 + \left[\frac{\Gamma_c}{2}(\omega-\omega_d) + \frac{\Gamma_d}{2}(\omega-\omega_c)\right]^2}. \tag{S-44}$$

As can be seen, Eq. (S-43a) takes the same form as Eq. (S-44), establishing a direct correspondence between the Green's function approach and the present formulation. Notably, the absorption spectrum obtained from the Zubarev Green's function formalism [1, 2, 7] is exactly reproduced by the total spectrum of the coupled system in the single-channel driving limit of our framework. This result further demonstrates that the conventional absorption spectrum emerges naturally as a limiting case of the present unified theory.

Taken together, these results establish a unified framework that not only recovers previously reported theories as limiting cases but also extends far beyond them. By encompassing a much broader range of driving configurations, the framework reveals diverse non-intrinsic spectral phenomena and enables on-demand manipulation of spectral responses and optical-field outputs. This universality establishes the framework as a broadly applicable and unifying paradigm for exploring and controlling non-Hermitian coupled systems across cavity quantum electrodynamics, nanophotonics, and beyond.

**Section 7. Ultra-narrow-linewidth in total spectral output of the coupled system.**

Under weak-coupling conditions, the total spectrum [Eq. (S-43b)] of the coupled system still exhibits linewidth compression relative to that of bare mode $d$ (Fig. S6). This indicates that the linewidth compression effect applies not only to individual channels but also to the total system response.

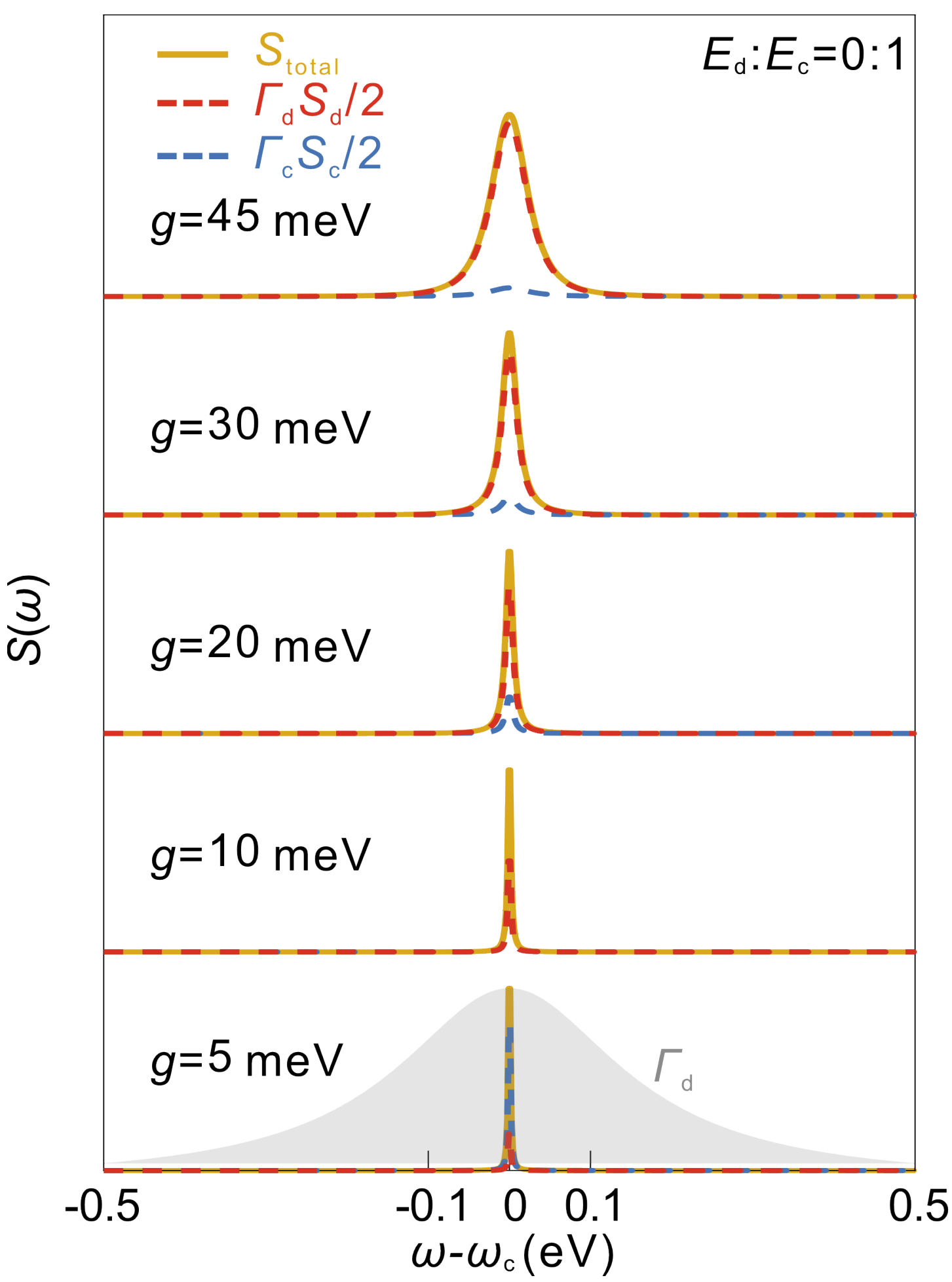

FIG. S6. The total normalized spectra of the coupled system (solid curves) under $E_d$: $E_c$ = 0:1 in the weak-coupling regime. The dashed curves represent the results of the spectra of each channel normalized to the maximum value of the total spectrum after applying the coefficient $A_d = \Gamma_d/2$ and $A_c = \Gamma_c/2$. The gray shaded area represents bare mode $d$. In calculations, $\Gamma_d$ = 200 meV and $\Gamma_c$ = 2 meV.

**Section 8. Derivation of spectral linewidths under exclusive Ch-*c* driving.**

In this section, we derive analytical expressions for the spectral linewidths (full widths at half maximum, FWHMs) of the Ch-*d*, Ch-*c*, and total spectral responses under resonant conditions when only low-loss mode *c* is driven, and analyze their dependence on coupling strength prior to the onset of spectral splitting (SS) at resonance ($\omega_d = \omega_c = \omega_0$).

The derivative of Eq. (S-34) is

$$\frac{dS_d^{\mathrm{dri},c}}{d\omega} = -g^2 \frac{4(\omega-\omega_0)\left[(\omega-\omega_0)^2 - g^2 - \frac{\Gamma_d\Gamma_c}{4}\right] + \frac{(\Gamma_d+\Gamma_c)^2}{2}(\omega-\omega_0)}{\left\{\left[(\omega-\omega_0)^2 - g^2 - \frac{\Gamma_d\Gamma_c}{4}\right]^2 + \frac{(\Gamma_d+\Gamma_c)^2}{4}(\omega-\omega_0)^2\right\}^2}. \tag{S-45}$$

Setting the derivative equal to zero, we obtain

$$2\left[(\omega-\omega_0)^2 - g^2 - \frac{\Gamma_d\Gamma_c}{4}\right] + \frac{(\Gamma_d+\Gamma_c)^2}{4} = 0, \tag{S-46}$$

Using Eq. (S-46), when $\omega = \omega_0$, the critical condition for SS can be obtained as

$$g_{d,\mathrm{SS}}^{\mathrm{dri},c} = \sqrt{\frac{\Gamma_c^2+\Gamma_d^2}{8}}, \tag{S-47}$$

Similarly, the derivative of Eq. (S-33) is

$$\frac{dS_c^{\mathrm{dri},c}}{d\omega} = (\omega-\omega_0)\frac{\left\{\begin{array}{l} 2\left[\left((\omega-\omega_0)^2 - g^2 - \frac{\Gamma_d\Gamma_c}{4}\right)^2 + \frac{(\Gamma_d+\Gamma_c)^2}{4}(\omega-\omega_0)^2\right] \\ -\left((\omega-\omega_0)^2 + \frac{\Gamma_d^2}{4}\right)\left[4\left((\omega-\omega_0)^2 - g^2 - \frac{\Gamma_d\Gamma_c}{4}\right) + \frac{(\Gamma_d+\Gamma_c)^2}{2}\right]\end{array}\right.}{\left[\left((\omega-\omega_0)^2 - g^2 - \frac{\Gamma_d\Gamma_c}{4}\right)^2 + \frac{(\Gamma_d+\Gamma_c)^2}{4}(\omega-\omega_0)^2\right]^2}. \tag{S-48}$$

As the derivative is equal to zero, $\omega$ satisfies

$$(\omega-\omega_0)^4 + \frac{\Gamma_d^2}{2}(\omega-\omega_0)^2 - \left[g^4 + \frac{1}{2}g^2\Gamma_d(\Gamma_c+\Gamma_d) - \frac{1}{16}\Gamma_d^4\right] = 0. \tag{S-49}$$

As $\omega = \omega_0$, the critical condition is given by

$$g_{c,\mathrm{SS}}^{\mathrm{dri},c} = \frac{1}{2}\sqrt{\Gamma_d\left[\sqrt{(\Gamma_c+\Gamma_d)^2+\Gamma_d^2} - (\Gamma_c+\Gamma_d)\right]}. \tag{S-50}$$

Then we calculate the variation of the FWHM with the coupling strength before spectral splitting occurs. When $\omega_d = \omega_c = \omega_0$, we can directly obtain the maximum value of the spectrum, which is achieved when $\omega = \omega_0$.

The maximum value of $S_d^{\mathrm{dri},c}$ is

$$\mathrm{Max}\left[S_d^{\mathrm{dri},c}\right]=\left|\frac{g}{g^2+\frac{\Gamma_c\Gamma_d}{4}}\right|^2. \tag{S-51}$$

The FWHM can be solved by the following equation

$$\left|\frac{g}{g^2+\frac{\Gamma_c\Gamma_d}{4}}\right|^2=2\left|\frac{g}{\left(\omega-\omega_0+i\frac{\Gamma_c}{2}\right)\left(\omega-\omega_0+i\frac{\Gamma_d}{2}\right)-g^2}\right|^2. \tag{S-52}$$

Solving it, we obtain

$$(\omega-\omega_0)^2=\frac{-M+\sqrt{M^2+4P^2}}{2}, \tag{S-53}$$

Where

$$M=\frac{\Gamma_d^2+\Gamma_c^2}{4}-2g^2, \tag{S-54}$$

$$P=g^2+\frac{\Gamma_c\Gamma_d}{4}. \tag{S-55}$$

Therefore

$$\mathrm{FWHM}\left[S_d^{\mathrm{dri},c}\right]=\sqrt{2\sqrt{M^2+4P^2}-2M}. \tag{S-56}$$

Similarly, the maximum value of $S_c^{\mathrm{dri},c}$ is

$$\mathrm{Max}\left[S_c^{\mathrm{dri},c}\right]=\left|\frac{\frac{\Gamma_d}{2}}{g^2+\frac{\Gamma_c\Gamma_d}{4}}\right|^2. \tag{S-57}$$

The FWHM can be solved from

$$\left|\frac{\frac{\Gamma_d}{2}}{g^2+\frac{\Gamma_c\Gamma_d}{4}}\right|^2=2\left|\frac{\omega-\omega_0+i\frac{\Gamma_d}{2}}{\left(\omega-\omega_0+i\frac{\Gamma_c}{2}\right)\left(\omega-\omega_0+i\frac{\Gamma_d}{2}\right)-g^2}\right|^2. \tag{S-58}$$

We can obtain

$$(\omega-\omega_0)^2=\frac{-(M-N)+\sqrt{(M-N)^2+4P^2}}{2}, \tag{S-59}$$

where

$$N = 8\left(\frac{g^2 + \frac{\Gamma_c \Gamma_d}{4}}{\Gamma_d}\right)^2, \tag{S-60}$$

Therefore

$$\mathrm{FWHM}\left[S_c^{\mathrm{dri},c}\right] = \sqrt{2\sqrt{(M-N)^2 + 4P^2} - 2(M-N)}. \tag{S-61}$$

As $g$ gradually decreases to 0, $\mathrm{FWHM}[S_c^{\mathrm{dri},c}]$ gradually approaches $\Gamma_c$ and the spectrum will degenerate into a standard Lorentzian profile ($S_{c0}$). For Ch-$d$, as $g = 0$, the spectrum cannot be obtained without mode coupling or external driving. Considering that $g \to 0$, neglecting $g^2$ in the denominator of Eq. (S-34), $S_d^{\mathrm{dri},c}$ can be written as

$$S_d^{\mathrm{dri},c} \approx g^2 \left|\frac{1}{\omega - \omega_0 + i\frac{\Gamma_c}{2}}\right|^2 \left|\frac{1}{\omega - \omega_0 + i\frac{\Gamma_d}{2}}\right|^2. \tag{S-62}$$

It can be seen that as $g \to 0$, the spectrum obtained from Ch-$d$ tends to be the spectrum of bare mode $c$ directly acting on that of $d$, with a FWHM of $\sqrt{\left(\sqrt{\Gamma_c^4 + \Gamma_d^4 + 6\Gamma_d^2\Gamma_c^2} - \Gamma_d^2 - \Gamma_c^2\right)/2}$ and a maximum intensity of $16g^2/{\Gamma_c}^2{\Gamma_d}^2$.

Besides, computing the derivative of the $S_{\mathrm{total}}^{\mathrm{dri},c}$ in Eq. (S-43b), we can obtain the critical value of the coupling strength for SS

$$g_{\mathrm{t,SS}}^{\mathrm{dri},c} = \frac{\Gamma_d}{2}\sqrt{\frac{\Gamma_d}{\Gamma_c + 2\Gamma_d}}, \tag{S-63}$$

Using the same method, the expression for the FWHM of the total spectrum before splitting occurs as a function of $g$ can be obtained:

$$\mathrm{FWHM}\left[S_{\mathrm{total}}^{\mathrm{dri},c}\right] = \sqrt{2\sqrt{(M-Q)^2 + 4P^2} - 2(M-Q)}. \tag{S-64}$$

where

$$Q = \frac{2\Gamma_c\left(\frac{\Gamma_d\Gamma_c}{4} + g^2\right)^2}{\Gamma_d g^2 + \frac{\Gamma_c\Gamma_d^2}{4}}. \tag{S-65}$$

**Section 9. Absence of linewidth compression under exclusive Ch-$d$ driving.**

In contrast to the exclusive driving of low-loss mode $c$, external driving applied solely to high-loss mode $d$ does not induce linewidth compression, but instead promotes the onset of spectral splitting (SS) in the self-excitation spectrum. In this case ($E_d$: $E_c$ = 1:0), the derivative of Eq. (S-31) at resonance is given by

$$\frac{dS_d^{\mathrm{dri},d}}{d\omega}=(\omega-\omega_0)\frac{\left\{\begin{array}{l}2\left[\left((\omega-\omega_0)^2-g^2-\frac{\Gamma_d\Gamma_c}{4}\right)^2+\frac{(\Gamma_d+\Gamma_c)^2}{4}(\omega-\omega_0)^2\right]\\-\left((\omega-\omega_0)^2+\frac{\Gamma_c^2}{4}\right)\left[4\left((\omega-\omega_0)^2-g^2-\frac{\Gamma_d\Gamma_c}{4}\right)+\frac{(\Gamma_d+\Gamma_c)^2}{2}\right]\end{array}\right.}{\left[\left((\omega-\omega_0)^2-g^2-\frac{\Gamma_d\Gamma_c}{4}\right)^2+\frac{(\Gamma_d+\Gamma_c)^2}{4}(\omega-\omega_0)^2\right]^2}. \quad \text{(S-66)}$$

Equating the derivative to zero, $\omega$ satisfies the relation

$$(\omega-\omega_0)^4+\frac{\Gamma_c^2}{2}(\omega-\omega_0)^2-\left[g^4+\frac{1}{2}g^2\Gamma_c(\Gamma_d+\Gamma_c)-\frac{1}{16}\Gamma_c^4\right]=0. \quad \text{(S-67)}$$

Thus, we can obtain the critical condition for SS

$$g_{d,SS}^{\mathrm{dri},d}=\frac{1}{2}\sqrt{\Gamma_c\left[\sqrt{(\Gamma_c+\Gamma_d)^2+\Gamma_c^2}-(\Gamma_c+\Gamma_d)\right]}. \quad \text{(S-68)}$$

From Eqs. (S-68) and (S-50), we can find that as $\Gamma_d > \Gamma_c$, $g_{c,SS}^{\mathrm{dri},c}$ is always more than $g_{d,SS}^{\mathrm{dri},d}$. Similar to Section 8, the FWHM of the output spectrum is

$$\mathrm{FWHM}\left[S_d^{\mathrm{dri},d}\right]=\sqrt{2\sqrt{(M-O)^2+4P^2}-2(M-O)}. \quad \text{(S-69)}$$

where $M$ and $P$ are shown in Eqs. (S-54) and (S-55), and

$$O=8\left(\frac{g^2+\frac{\Gamma_c\Gamma_d}{4}}{\Gamma_c}\right)^2. \quad \text{(S-70)}$$

The situation for the coupling-excitation is identical to that in Eqs. (S-32) and (S-34).

$$g_{d,\mathrm{SS}}^{\mathrm{dri},c}=g_{c,\mathrm{SS}}^{\mathrm{dri},d}. \quad \text{(S-71)}$$

$$\mathrm{FWHM}\left[S_c^{\mathrm{dri},d}\right]=\mathrm{FWHM}\left[S_d^{\mathrm{dri},c}\right]. \quad \text{(S-72)}$$

In the case where mode $d$ is individually driven, the SS threshold of the total spectrum is evaluated as

$$g_{\mathrm{t,SS}}^{\mathrm{dri},d}=\frac{\Gamma_c}{2}\sqrt{\frac{\Gamma_c}{\Gamma_d+2\Gamma_c}}. \quad \text{(S-73)}$$

The expression for the FWHM of the total spectrum when mode $d$ is excited before SS

occurs as a function of $g$ can be obtained:

$$\mathrm{FWHM}\left[S_{\mathrm{total}}^{\mathrm{dri},d}\right]=\sqrt{2\sqrt{(M-R)^2+4P^2}-2(M-R)}. \tag{S-74}$$

where

$$R=\frac{2\Gamma_d\left(\frac{\Gamma_d\Gamma_c}{4}+g^2\right)^2}{\Gamma_c g^2+\frac{\Gamma_d\Gamma_c^2}{4}}. \tag{S-75}$$

In Fig. S7(a), we plot the normalized spectra for different coupling strengths under exclusive driving of the high-dissipation mode $d$. It can be observed that when only high-dissipation mode $d$ is driven, the self-excitation spectra exhibit pronounced SS behavior, even at very weak coupling strengths, whereas the coupling-excitation spectra undergo linewidth broadening while maintaining a single peak, implying that a higher coupling strength is required to satisfy the critical condition for SS of Ch-$c$. In Fig. S7(b), we further plot the linewidths of the self-excitation spectra and the coupling-excitation spectra prior to splitting as a function of $g$ for both single-driving scenarios. The results demonstrate that exclusive driving of the high-dissipation mode $d$ suppresses the convergence effect and yields the lowest SS threshold for Ch-$d$ (dashed red curve), which is lower than that of Ch-$c$ under the same driving condition (dashed purple curve) and also lower than both Ch-$d$ (dashed purple curve) and Ch-$c$ (solid blue curve) under exclusive driving of the low-dissipation mode $c$.

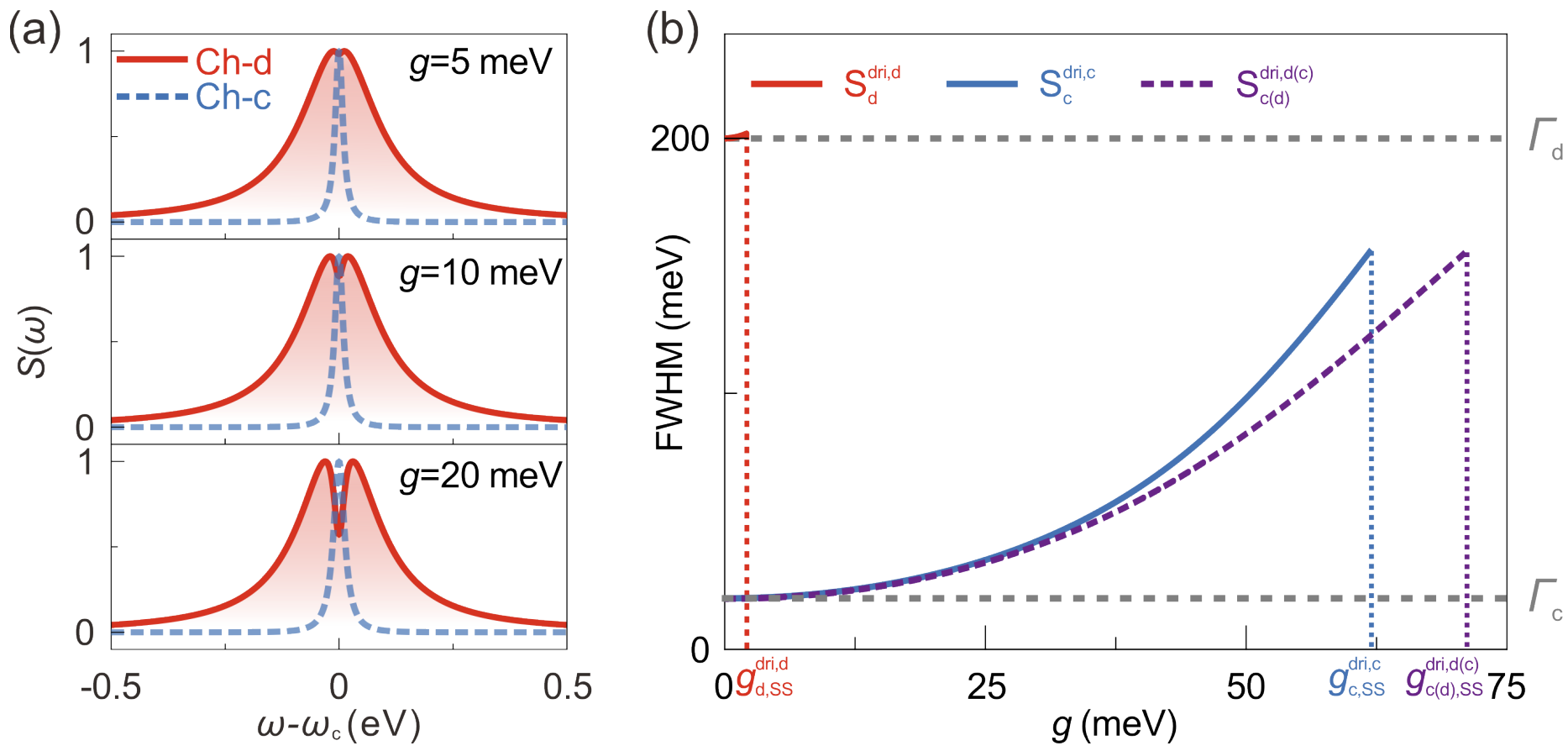

FIG. S7. (a) Normalized output spectra from Ch-$d$ and Ch-$c$, corresponding to the driving conditions $E_d$: $E_c$ = 1:0 with $g$ = 5, 10, and 20 meV, respectively. (b) The FWHMs of the $S_d^{\mathrm{dri},d}$, $S_c^{\mathrm{dri},c}$ and $S_{c(d)}^{\mathrm{dri},d(c)}$ as a function of $g$. In calculations, $\Gamma_d$ = 200 meV and $\Gamma_c$ = 20 meV.

**Section 10. Spectral Rabi splitting (SRS) disappearance.**

**10.1 Interference-induced SRS disappearance in Ch-*d* and Ch-*c*.**

As shown in the main text, balanced driving configurations ($E_d : E_c$ = 1:1) leads to the disappearance of the SRS feature in both Ch-*d* and Ch-*c* under strong coupling. To reveal the origin of this effect, we expand Eqs. (S-17) and (S-18) and decompose the spectral intensities into self-excitation, coupling-excitation, and the interference contributions

$$\begin{aligned} S_d &\propto | E_d\psi_d + E_c\psi_g |^2 \\ &= | E_d\psi_d |^2 + | E_c\psi_g |^2 + 2E_dE_c\mathrm{Re}\left(\psi_d^*\psi_g\right), \\ &= I_d^{\mathrm{SE}} + I_d^{\mathrm{CE}} + I_d^{\mathrm{INT}} \end{aligned} \tag{S-76}$$

$$\begin{aligned} S_c &\propto | E_c\psi_c + E_d\psi_g |^2 \\ &= | E_c\psi_c |^2 + | E_d\psi_g |^2 + 2E_dE_c\mathrm{Re}\left(\psi_c^*\psi_g\right), \\ &= I_c^{\mathrm{SE}} + I_c^{\mathrm{CE}} + I_c^{\mathrm{INT}} \end{aligned} \tag{S-77}$$

where $I_{d(c)}^{\mathrm{SE}}$, $I_{d(c)}^{\mathrm{CE}}$ and $I_{d(c)}^{\mathrm{INT}}$ denote the self-excitation, coupling-excitation, and interference contributions to the spectral intensity in the Ch-*d* and Ch-*c*, respectively. Their explicit expressions are

$$I_{d(c)}^{\mathrm{SE}} = \frac{| E_{d(c)} |^2 \left[ (\omega-\omega_{c(d)})^2 + \frac{\Gamma_{c(d)}^2}{4} \right]}{\left[ (\omega-\omega_d)(\omega-\omega_c) - \frac{\Gamma_d\Gamma_c}{4} - g^2 \right]^2 + \left[ (\omega-\omega_d)\frac{\Gamma_c}{2} + (\omega-\omega_c)\frac{\Gamma_d}{2} \right]^2}, \tag{S-78}$$

$$I_{d(c)}^{\mathrm{CE}} = \frac{| E_{c(d)} |^2 g^2}{\left[ (\omega-\omega_d)(\omega-\omega_c) - \frac{\Gamma_d\Gamma_c}{4} - g^2 \right]^2 + \left[ (\omega-\omega_d)\frac{\Gamma_c}{2} + (\omega-\omega_c)\frac{\Gamma_d}{2} \right]^2}, \tag{S-79}$$

$$I_{d(c)}^{\mathrm{INT}} = \frac{2E_dE_cg(\omega-\omega_{c(d)})}{\left[ (\omega-\omega_d)(\omega-\omega_c) - \frac{\Gamma_d\Gamma_c}{4} - g^2 \right]^2 + \left[ (\omega-\omega_d)\frac{\Gamma_c}{2} + (\omega-\omega_c)\frac{\Gamma_d}{2} \right]^2}. \tag{S-80}$$

Under simultaneous two-channel balanced driving, the spectral intensity in each channel can be decomposed into self-excitation, coupling-excitation, and interference contributions. As shown in Figs. S8(a) and S8(b), both the self-excitation and coupling-excitation terms retain a clear two-branch structure under strong coupling, indicating that the underlying mode splitting remains intact. In contrast, the interference contribution exhibits a highly asymmetric spectral distribution and selectively suppresses one of the two split branches. As a result, destructive interference nearly cancels the spectral intensity of one branch, while constructively enhancing the other. The total spectrum therefore evolves from a split two-peak profile into an apparently single-branch response, leading to the disappearance of the SRS feature. These results demonstrate that the observed SRS disappearance does not originate from the collapse of the strong-coupling-induced mode splitting itself, but rather from interference between the self-excitation and coupling-excitation pathways. Notably, the split spectral branches remain clearly resolved in both the self-excitation and coupling-excitation contributions, confirming that the system remains in the strong-coupling regime even when the splitting becomes invisible in the measured spectrum. The disappearance of SRS therefore reflects a spectroscopic consequence of interference-induced spectral redistribution, rather than a breakdown of the underlying strong-coupling state.

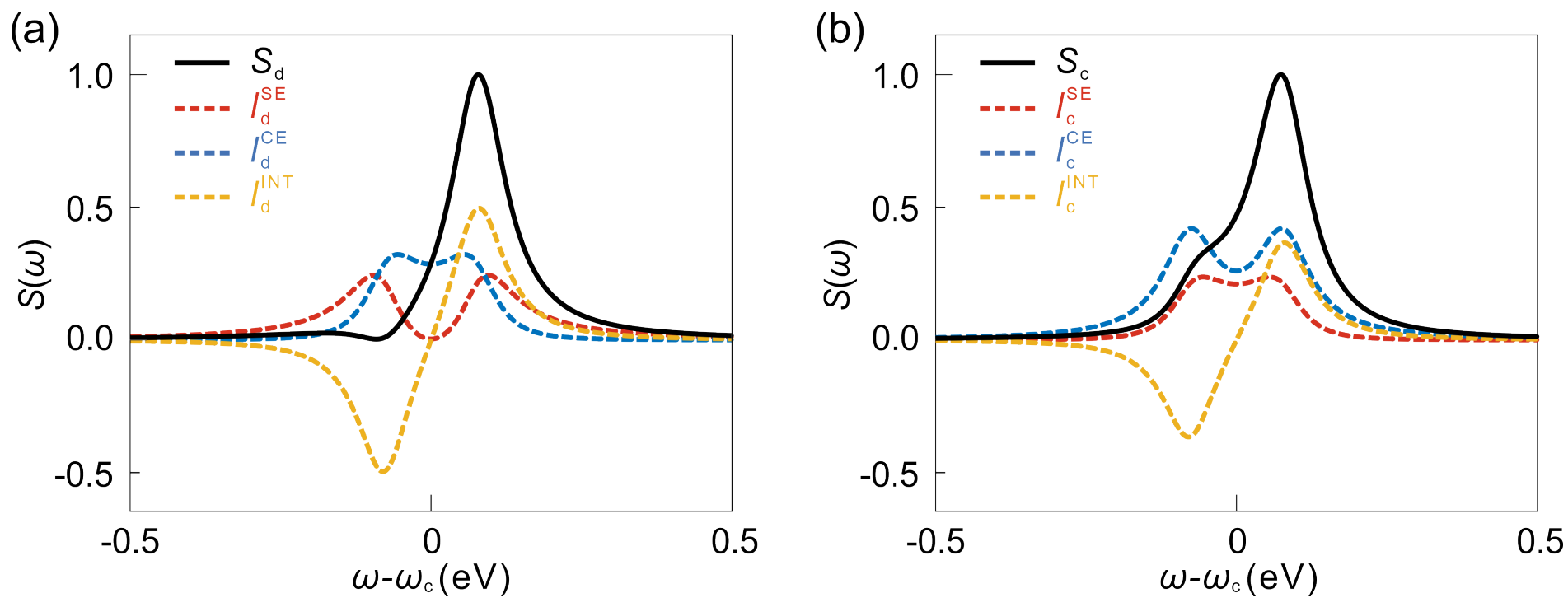

FIG. S8. Spectral decoupling of Ch-$d$ (a) and Ch-$c$ (b) under $E_d : E_c$ = 1:1 and $g$ = 90 meV. In calculations, $\Gamma_d$ = 200 meV and $\Gamma_c$ = 20 meV.

### 10.2 SRS disappearance in the total spectrum.

The interference-induced spectral redistribution is not limited to the individual detection channels but also strongly influences the total spectral response, which is constructed from the weighted superposition of the Ch-$d$ and Ch-$c$ spectra ($S_{\mathrm{total}} = A_d S_d + A_c S_c$). Consequently, interference continues to play a decisive role in shaping the observed spectral line-shape of the coupled system. In Fig. S9, we plot the normalized total spectra of the coupled system with $\Gamma_d$ fixed and $\Gamma_c$ varied, while the spectra of Ch-$d$ and Ch-$c$ are shown as dashed curves for comparison. Similar to the behavior observed in the individual channels, the SRS feature also disappears in the total spectrum and is replaced by an apparently single-peak line-shape. Notably, the same disappearance of SRS is observed in both individual channels and the total spectrum, demonstrating that the effect is not channel-specific but is governed by the same interference mechanism across all spectral responses.

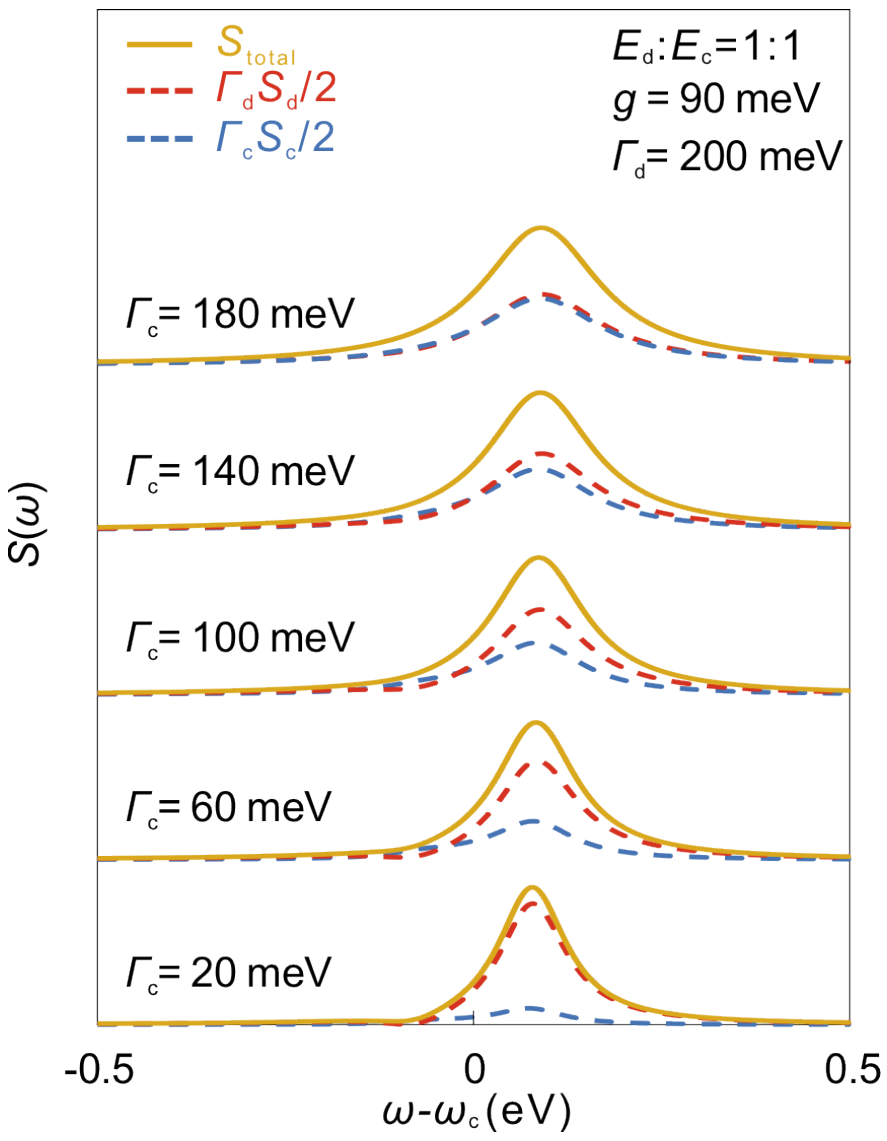

FIG. S9. The total normalized spectra of the coupled system (solid curves) for $E_d$: $E_c$ = 1:1 at fixed $\Gamma_d = 200$ meV and different $\Gamma_c$. The dashed curves represent the results of the spectra of each channel normalized to the maximum value of the total spectrum after accounting for the coefficient $A_d = \Gamma_d/2$ and $A_c = \Gamma_c/2$.

### 10.3 Giant frequency shift with fixed linewidth.

When the coupled system is under perfectly balanced conditions ($E_d$: $E_c = 1$: 1, $\Gamma_d$: $\Gamma_c = 1$: 1, and $\omega_d = \omega_c$), the parameters can be written as $E_d = E_c = E_0$, $\Gamma_d = \Gamma_c = \Gamma_0$, and $\omega_d = \omega_c = \omega_0$. Eqs. (S-15) and (S-16) then reduce to

$$d = c = E_0 \frac{\left(\omega - \omega_0 + i\frac{\Gamma_0}{2}\right) + g}{\left(\omega - \omega_0 + i\frac{\Gamma_0}{2}\right)\left(\omega - \omega_0 + i\frac{\Gamma_0}{2}\right) - g^2} . \tag{S-81}$$

The denominator is expanded using the difference of squares formula, and we can obtain

$$d = c = E_0 \frac{1}{\omega - (\omega_0 + g) + i\frac{\Gamma_0}{2}} . \tag{S-82}$$

From Eq. (S-82), the two modes become fully degenerate, and the spectra of Ch-*d* and Ch-*c* reduce to the same expression:

$$S_d = S_c \propto \left| \frac{1}{\omega - (\omega_0 + g) + i\frac{\Gamma_0}{2}} \right|^2 . \tag{S-83}$$

From the above expression, the spectrum is characterized by a Lorentzian profile with central frequency $\omega_0 + g$ and fixed linewidth $\Gamma_0$. Notably, the central frequency of this Lorentzian spectrum shifts linearly with increasing coupling strength $g$ without saturation, whereas the linewidth remains unchanged, as shown in Fig. S10(a). The spectrum remains a standard single-peak Lorentzian line-shape, with no spectral Rabi splitting, accompanied by an unsaturated giant frequency shift and a fixed linewidth.

In fact, from the denominator of Eq. (S-83), it follows that increasing $g$ induces only a resonance frequency shift without modifying the linewidth. This results in a giant unsaturated spectral shift under perfectly balanced conditions. Figure S10(b) further confirm this behavior in the smaller-linewidth regime. These results highlight the potential of this mechanism for tunable light sources and quantum photonic applications.

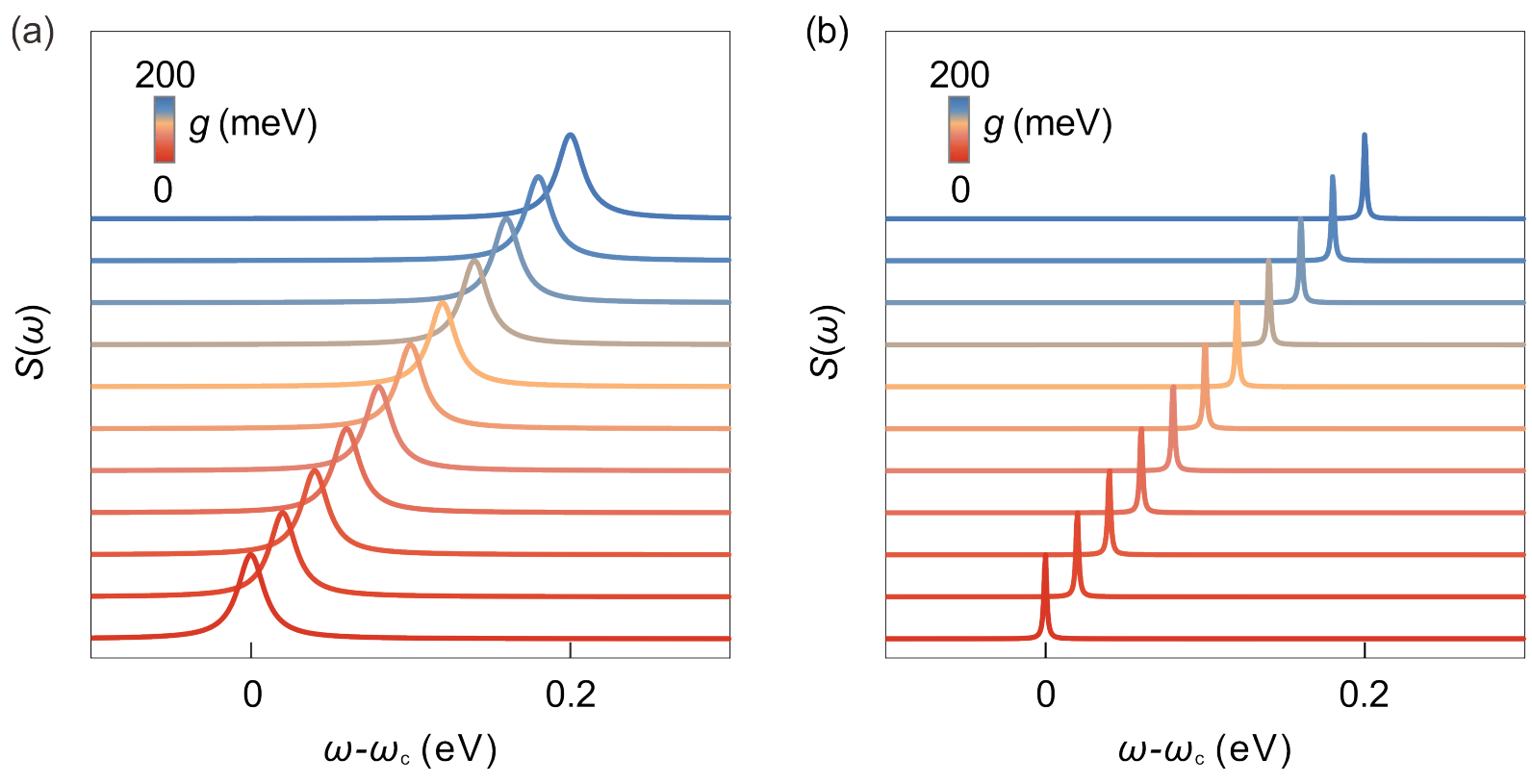

FIG. S10. Spectral response under the condition of (a) $\Gamma_d = \Gamma_c = 20$ meV and (b) $\Gamma_d = \Gamma_c = 2$ meV with $E_d : E_c = 1{:}1$.

**Section 11. Analytical derivation of SRS under single-channel driving.**

We investigate the SRS of the coupling-excitation spectrum under single-mode driving. For individual mode excitation, the splitting width $\Delta_{\mathrm{SRS}.g}$ is obtained from the derivative of the $S_{d(c)}^{\mathrm{dri},c(d)}$, when $\omega_d = \omega_c = \omega_0$. According to Eq. (S-46), we can obtain

$$\omega_{\pm} = \omega_0 \pm \sqrt{g^2 - \frac{\Gamma_c^2 + \Gamma_d^2}{8}}. \tag{S-84}$$

Then, the spectral splitting obtained from Ch-$d(c)$ ($\Delta_{\mathrm{SRS},g}$) (when $g$ satisfies the splitting condition) is given by

$$\Delta_{\mathrm{SRS},g} = \omega_+ - \omega_- = 2\sqrt{g^2 - \frac{\Gamma_c^2 + \Gamma_d^2}{8}}. \tag{S-85}$$

**Section 12. Simulations.**

The simulations employed a two-step background-scattered field formulation. First, the background electromagnetic field distribution of the bare Ag NAC was calculated under normal incidence with X-polarized plane waves, applying periodic boundary conditions in both the X and Y directions. The port boundary conditions were set in the Z-direction. Subsequently, the Au NR was introduced into this background field, and a 2 nm truncation was applied to the bottom of the nanorod to account for practical experimental conditions. To suppress unphysical boundary reflections, the entire computational domain was enclosed by perfectly matched layers (PMLs). The absorption power was extracted by integrating the power loss density over the volume of the Au NR. For the simulations, the refractive index of $SiO_2$ was set to 1.5 and the ambient refractive index was 1. The dielectric functions of Au and Ag are obtained from Refs. [8] and [9], respectively.

From the inset of Fig. 6(c) in the main text, the coupling between Au NR and Ag NAC is mainly located at their interface. We now extract the $|E_x|$ distribution of the X-Z cross-section at the position of $\delta_y$ = 220 nm in the uncoupled Ag NAC [Fig. S11(a)], and the intensity variation of $|E_x|$ on the $SiO_2$ surface of the Ag NAC can be clearly observed. We further plotted the variation of $E_x$ intensity on the surface with position. According to the fitting parameters of the simulation results in Fig. 6(c), both the excitation intensity ratio of mode *c* and the coupling strength of the system are positively correlated with the intensity of $|E_x|$. This reveals the physical mechanism by which altering the position of the Au NR within the cavity can tune its absorption power spectrum.

For further verification, we extracted the $|E_x|$ distribution of the Y-Z cross-section at the position of $\delta_x$ = 200 nm, as shown in Fig. S11(b). It can be observed that the $|E_x|$ intensity on the $SiO_2$ surface of the cavity is uniform. According to the analysis above, when the $\delta_x$ is fixed at 200 nm, the absorption spectrum of the Au NR will remain consistent with varying $\delta_y$. Figure S11(c) presents the simulated normalized absorption cross sections of Au NR with varying $\delta_y$ at $\delta_x$ = 200 nm, and the spectral lineshapes maintain high consistency, exactly as predicted. This also demonstrates that our proposed classical system enables not only diverse spectral modulation, but also highly robust linewidth narrowing.

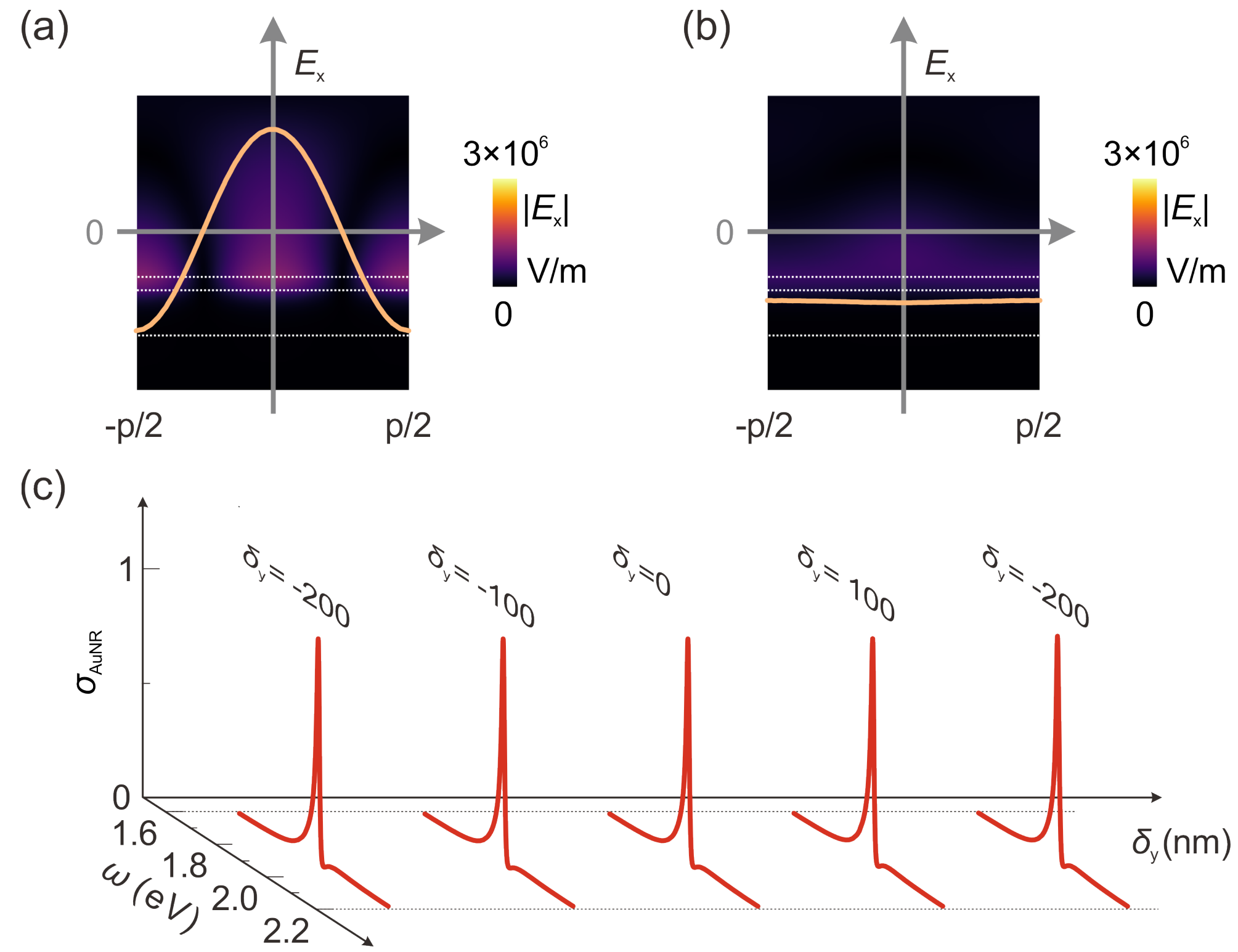

FIG. S11. (a) $|E_x|$ distributions in the X-Z cross-section at $\delta_y$ = 220 nm. (b) $|E_x|$ distributions in the Y-Z cross-section at $\delta_x$ = 200 nm. The yellow curves in (a) and (b) indicate the variation of the $E_x$ intensity with position at the top $SiO_2$ interface of the Ag NAC. (c) Evolution of the normalized absorption cross sections of the Au NR at various lateral displacements $\delta_y$ with fixed $\delta_x$ = 200 nm.

We additionally assess the contribution to the far-field scattering in forming the bare Ag NAC mode through multipole expansion in Cartesian coordinates. The decomposed far-field scattering powers of multipoles are calculated based on current density.

Electric dipole:

$$I_P = \frac{2\omega^4}{3c^3} \left| \frac{1}{i\omega} \int \vec{j} d^3 r \right|^2 \tag{S-86}$$

Magnetic dipole:

$$I_M = \frac{2\omega^6}{3c^5} \left| \frac{1}{2c} \int (\vec{r} \times \vec{j}) d^3 r \right|^2 \tag{S-87}$$

Electric quadrupole:

$$I_Q^{(e)} = \frac{\omega^6}{5c^5} \sum \left| \frac{1}{2i\omega} \int \left[ r_\beta j_\alpha + r_\alpha j_\beta - \frac{2}{3} (\vec{r} \cdot \vec{j}) \delta_{\beta\alpha} \right] d^3 r \right|^2 \tag{S-88}$$

Magnetic quadrupole:

$$I_Q^{(m)} = \frac{\omega^6}{20c^5}\sum\left|\frac{1}{3c}\int\left[(\vec{r}\times\vec{j})_\beta r_\alpha + (\vec{r}\times\vec{j})_\alpha r_\beta\right]d^3r\right|^2 \tag{S-89}$$

Toroidal dipole:

$$I_T = \frac{2\omega^6}{3c^5}\left|\frac{1}{10c}\int\left[(\vec{r}\cdot\vec{j})\vec{r} - 2r^2\vec{j}\right]d^3r\right|^2 \tag{S-90}$$

It can be seen from Fig. S12 that the mode of Ag NAC is mainly provided by the resonance of the electric dipole.

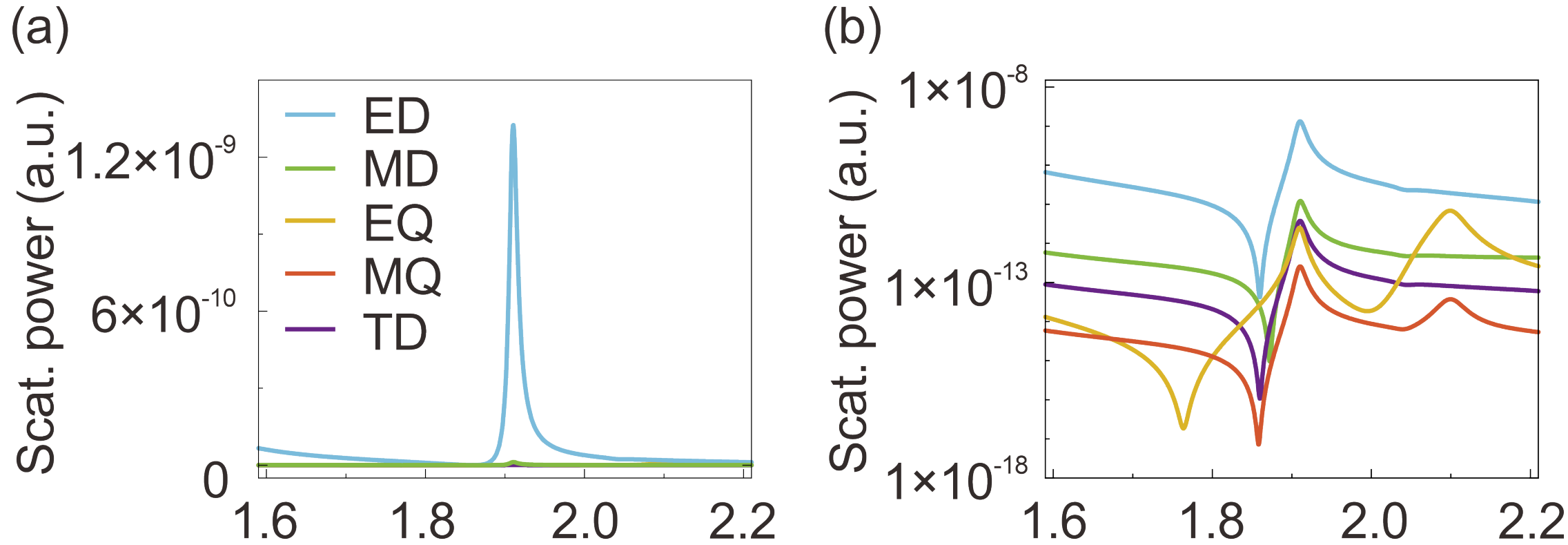

FIG. S12. The calculated powers scattered by different multipoles of the Ag NAC in (a) linear and (b) logarithmic coordinates.